\documentclass[prx, twocolumn, superscriptaddress]{revtex4-2}

\usepackage[utf8]{inputenc}

\usepackage{graphicx}
\usepackage{amsmath}
\usepackage{nameref,hyperref}
\hypersetup{
    colorlinks=true,
    citecolor=red,
    }

\usepackage[aboveskip=1pt,labelfont=bf,labelsep=period,justification=raggedright,singlelinecheck=off]{caption}

\newcommand{\fig}{Fig.~}
\newcommand{\supp}{Supplementary~}
\newcommand{\M}{{M}} 

\newcommand{\mat}[1]{\boldsymbol{\mathbf{{#1}}}}

\newcommand{\uni}{Department of Computer Sciences, University of T\"ubingen, Germany}
\newcommand{\mpi}{Max Planck Institute for Biological Cybernetics, T\"ubingen, Germany}
\newcommand{\boc}{Department of Computing Sciences, Bocconi University, 20136 Milano, Italy}
\newcommand{\hih}{Department of Neurology and Epileptology, University of Tübingen, 72076 Tübingen, Germany}
\newcommand{\imp}{Department of Bioengineering, Imperial College London, SW7 2AZ London}

\begin{document}

\title{Cell-type-specific modularity in E/I networks shapes dynamical regimes and computational capabilities}

\author{Emmanouil Giannakakis} 
\thanks{These authors contributed equally.}
\affiliation{\uni}
\affiliation{\mpi}
\affiliation{\imp}

\author{Victor Buend\'ia}
\thanks{These authors contributed equally.}
\affiliation{\uni}
\affiliation{\mpi}
\affiliation{\boc}

\author{Oleg Vinogradov}
\affiliation{\uni}
\affiliation{\mpi}
\affiliation{\hih}

\author{Anna Levina}
\affiliation{\uni}
\affiliation{\mpi}

\begin{abstract}
Structural modularity is a ubiquitous feature of brain networks that shapes neural dynamics and information processing. Yet excitatory and inhibitory populations can differ in their modular organization, and it remains unclear how this asymmetry affects network dynamics and computation.
Here, using dynamical mean-field theory and simulations of networks performing complex temporal tasks, we show that independently tuning excitatory and inhibitory modularity controls the dynamical regimes of balanced E/I networks. In particular, such tuning can generate a dominant spectral outlier despite overall E/I balance, producing a topology-controlled transition between bulk- and low-rank-dominated regimes, with different dynamical bifurcations leading to chaos and bistability, respectively.
In chaotic time-series prediction tasks, computational performance is enhanced near these dynamical boundaries, including both the onset of chaos and the transition to bistability. Favorable configurations exhibit strongly modular excitation and comparatively homogeneous inhibition, similar to what is observed in cortical connectivity.
\end{abstract}

\maketitle

\section{Introduction}

\begin{figure*}
  \includegraphics[width=\linewidth]{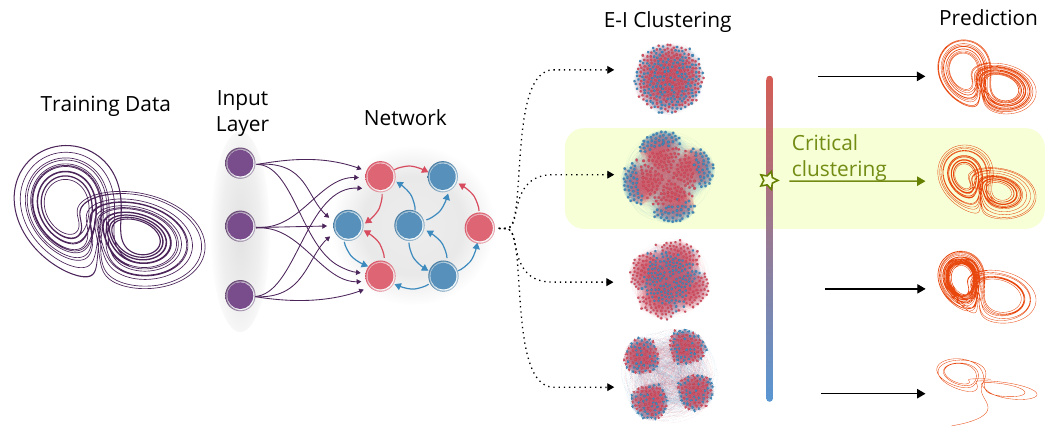}
  \caption{\textbf{Different levels of $E$ and $I$ modularity control network dynamics and computational capabilities.} We investigate the impact of synapse-type-specific recurrent connectivity on the dynamics of a balanced network and evaluate its impact on the network's computational capabilities. Using dynamic mean-field theory, we map different network topologies to distinct dynamical states and, using a reservoir computing framework, we show how the analytically-derived dynamical state controls the network's performance in a time series prediction task.}
 \label{fig:clustering_explanation}
\end{figure*}

Connectivity in biological neural networks is highly structured and inhomogeneous, often organized into modules or clusters with dense intra-module and sparse inter-module connectivity \cite{hilgetag2004clustered, valencia2009complex, sporns2016modular}.
 Modular and clustered connectivity structures have been shown to significantly impact network dynamics, for example, by controlling population bursting \cite{blanco2024population}, generating cascading dynamics \cite{galstyan2007cascading}, and increasing the robustness of attractor dynamics \cite{pradhan2011modular}. Beyond this, modular network connectivity has been linked to cortical-like activity patterns \cite{yamamoto2018impact, lundqvist2006attractor} and shown to impact information processing \cite{houk1994information} and task performance \cite{kirsch2018modular}.

At the same time, biological neural networks are comprised of diverse neuron types that follow distinct and highly specific connectivity patterns \cite{schneider2023cell}.  In particular, excitatory and inhibitory neurons differ not only in their functional roles but also in their local and long-range connectivity, with various inhibitory subtypes forming heterogeneous and spatially structured networks \cite{levy2012spatial, znamenskiy2024functional}. This cell-type specificity of wiring suggests that modular organization may differ between excitation and inhibition.

Despite this biological complexity, most theoretical studies of modular and clustered brain networks have focused on excitatory populations alone \cite{LitwinKumar2012natneuro, Setareh2017frontiers, yamamoto2018impact}. Models investigating interplay between excitation and inhibition have largely focused on the dynamics of balanced networks under spatial constraints \cite{Rosenbaum2014PRX, Rosenbaum2017NatNeuro}. While recent work has demonstrated that clustered E/I architectures can substantially influence population dynamics \cite{Rost2018BiologicalCybernetics, Rostami2020biorxiv, clark_structure_2025}, the impact of independently tunable, cell-type–specific modular organization of excitation and inhibition on network dynamics and information processing has not been studied.

Notably, variations in excitatory clustering alone, combined with uniform inhibition, are sufficient to induce rich dynamical regimes characterized by long autocorrelation timescales \cite{LitwinKumar2012natneuro}. Therefore, different $E$ and $I$ modular structures could have serious implications on the computational capabilities of neuronal networks. 
Consistent with this idea, clustered spiking E/I networks have been shown to enhance learning efficiency in reinforcement learning settings \cite{Weidel2021frontiers}.

Here we study the dynamics and computational properties of a modular, balanced rate network with cell-type–specific connectivity. Using dynamic mean-field theory, we characterize how independent tuning of excitatory and inhibitory modularity shapes the network’s dynamical regime. We then relate these dynamical states to computational performance on a complex time-series prediction task using an echo state network framework, and examine how structured external inputs interact with E/I-modular recurrent connectivity. Together, this approach links cell-type–specific network architecture, emergent population dynamics, and computational performance within a unified theoretical framework.

\section{Results}

\begin{figure*}
  \includegraphics[width=\linewidth]{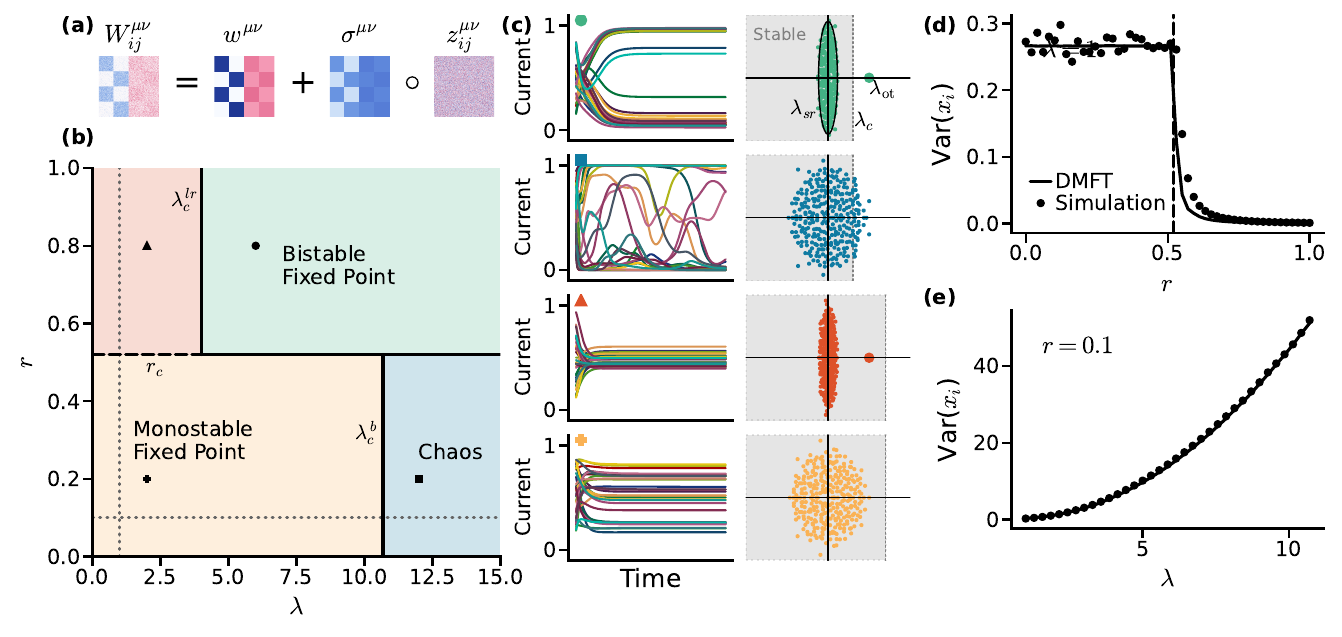}
  \caption{\textbf{Dynamics of the reservoir network.} \textbf{a.} Visual decomposition of the $N\times N$ connectivity matrix as mean and variance components. \textbf{b.} Phase diagram of the recurrent network in the absence of external input, as predicted analytically by the dynamic mean-field theory. Solid lines represent bifurcations where the stability of fixed points changes. The dashed line indicates a transition leading to an abrupt change in the variance around the monostable fixed point. The dotted gray horizontal and vertical lines indicate the sections shown in panels (d) and (e). \textbf{c. }Example trajectories of individual for networks of $N=3000$ for all 4 different phases of the diagram (left) and the corresponding spectrum of the connectivity matrix (right). The theoretical prediction of the outlier eigenvalue is shown by a larger dot. The dotted gray lines indicate the threshold of instability ($\lambda_c ^b$ or $\lambda_c ^{lr}$) and the gray background highlights the region of stability. 
  \textbf{d.} Variance of the distribution of fixed points around their mean as a function of the clustering for a fixed spectral radius $\lambda=1$. A continuous line indicates the theoretical results from DMFT, while dots are values obtained by averaging 10 networks of size $N=1000$. \textbf{e.} Same as (d), as a function of the spectral radius with fixed $r=0.1$.}
   \label{fig:dynamics}
\end{figure*}

We use a network of ${N}$ rate neurons split into ${M}$ modules. 
Each module contains an equal number of excitatory and inhibitory neurons, which respectively form excitatory and inhibitory populations. Thus, the network has $2 {M}$ populations, indexed by Greek letters $\mu \in \{1,\dots, 2\M\}$. Populations $\mu \in \{1,\ldots,M\}$ correspond to excitatory ones and $\mu \in \{M+1, \ldots, 2M\}$ to inhibitory ones.  We choose the number of neurons $N$ to be a multiple of $2M$, so each population has exactly $N/(2M)$ neurons.
We vary the network’s modularity by redistributing connection density between intra-module and inter-module links while keeping the total connection density fixed. 
To this end, we introduce a modularity parameter $r \in [0, 1]$, where $r = 0$ corresponds to uniform connectivity without modular structure and $r=1$ to fully disconnected modules. Specifically, the probability of connection from neuron $j$ in population $\nu$ to neuron $i$ in population $\mu$ is given by:
\begin{equation}
p_{ij}^{\mu \nu} = 
   \begin{cases}
\dfrac{p}{\M} \cdot \left[1 + (\M-1) \cdot r_{J(\mu)J(\nu)}  \right], \ I(\mu) = I(\nu) \\[6pt]
\dfrac{p}{\M} \cdot \left[  1 - r_{J(\mu)J(\nu)}  \right] ,  \ \textrm{otherwise}
\end{cases}
\label{eq:probconn}
\end{equation}
where $p$ sets the overall connection density, the function $J(\mu) \in \{ E,I\}$ identifies the cell type, with $J(\mu)=E$ for $\mu \leq M$ and $J(\mu)=I$ if $\mu>M$. $I(\mu)=\mu \mod M$ denotes the module index.
This parametrization preserves the expected in-degree of each population while systematically biasing connections toward intra-module or inter-module wiring. 
For all realized connections, synaptic weights are drawn independently from a prescribed distribution (see Methods).
Here, we study the effects of different modularity levels for the excitatory and inhibitory populations by varying $r_E = r_{IE}= r_{EE}$ and $r_I = r_{II}= r_{EI} $, thus allowing the modularity to depend only on the identity of the presynaptic neuron. This construction allows us to control the fraction of excitatory and inhibitory inputs each neuron receives from its module relative to the total number of incoming excitatory and inhibitory connections (\fig\ref{fig:clustering_explanation}, \fig\ref{fig:dynamics} a). Thus, the modularity parameters $r_E,\, r_I$ act as explicit control parameters of the network topology, enabling systematic tuning of the connectivity structure.

\subsection{The level of excitatory vs inhibitory modularity controls the network's dynamics}

\newcommand{\unw}{\mat{\tilde W}}

We begin by studying the dynamics of E/I modular networks in the absence of external inputs.
Specifically, we investigate the distribution of stable fixed points and their stability.  
In recurrent networks used for computation, these intrinsic dynamics are known to strongly influence computational performance.
In particular, for echo-state networks, the echo-state property requires that perturbations decay over time, which corresponds to the largest Lyapunov exponent being negative \cite{yildiz2012NN, lukosevicius2012NNTotTSE, caluwaerts20132IJCNNI}.
At the same time, performance is often maximized near the \emph{edge of chaos}, where the largest Lyapunov exponent approaches zero from below \cite{bertschinger2004real, lukosevicius2012NNTotTSE, jiang2019PRR}.
Identifying the location in parameter space of the transition from fixed points to chaos is therefore of central importance.

In a random heterogeneous network, the transition from a stable fixed point to chaotic dynamics is controlled by the spectral properties of the connectivity matrix.
When a dense bulk of eigenvalues crosses the stability threshold, many eigenmodes destabilize simultaneously, leading to chaotic dynamics. 
In contrast, when a single eigenvalue crosses the threshold, only one mode is destabilized, resulting in a transition to a new stable fixed point or limit cycle. Recent work by Dahmen \emph{et al.} has emphasized the qualitative differences between these two types of transitions and their impact on the correlation structure of neurons \cite{dahmen2019PNAS}.

We therefore begin by determining the stability of fixed points in our system, constructing its full phase diagram.
These theoretical results form the basis for our later analysis of reservoir performance.
We summarize the main results and their implications here; we present the full analytical derivation step by step in \supp~I-III.

We describe the recurrent connectivity in the network by a $N \times N$ block-structured matrix, where each block is approximated by random Gaussian weights with block-specific mean and variance (\fig\ref{fig:dynamics}a). The approximation of connectivity by a random Gaussian block matrix has proven to accurately predict network dynamics in similar settings \cite{kadmon2015PRX, aljadeff2015PRL, mastrogiuseppe2017PCB, mastrogiuseppe2018N, clark_structure_2025}. Within this framework, the elements of the (unnormalized) connectivity matrix $\unw$ can be written as 
\begin{equation}
    \tilde W_{ij} ^{\mu \nu} \simeq w_{\mu \nu}  + \sigma_{\mu \nu}  z_{ij} ^{\mu \nu}, \label{eq:decomposition}
\end{equation}
where $z_{ij}^{\mu \nu} \sim \mathcal{N}(0,1)$ are independent standard Gaussian random variables, and the reduced matrix elements $w_{\mu \nu}$ and $\sigma_{\mu \nu}$ denote the mean and standard deviation of the weights between populations $\mu$ and $\nu$.
The reduced matrices $\mat{w}$ and $\mat{\sigma}$ are of size $2M \times 2M$, with each entry corresponding to a specific population pair.
Indices $i,j \in \{1,2,\ldots, N/(2M)\}$ label individual neurons within each population block.

We denote the largest eigenvalue of $\unw$ by $\lambda_{\tilde W}$. The spectrum of $\unw$ scales with system size, so normalization is required to take the thermodynamic limit $N\to+\infty$. In our setup, we normalize the connectivity matrix $\mat{W}$ of our echo-state networks to have largest eigenvalue $\lambda$, which we treat as a control parameter for the reservoir. This normalization differs from the one typically used in the study of dynamical mean-field recurrent neural networks \cite{kadmon2015PRX, aljadeff2015PRL, mastrogiuseppe2017PCB}, but it yields qualitatively similar results (see \supp IIIB for a detailed comparison). Thus,
\begin{equation}
    \mat{W} = \frac{\lambda}{\lambda_{\tilde W}} \unw.
\end{equation}

The spectrum of the full connectivity matrix $\mat{W}$ follows directly from $\unw$. The spectrum of $\unw$ is determined by the spectra of the reduced mean and variance matrices, $\mat{w}$ and $\mat{\sigma}$ \cite{mastrogiuseppe2017PCB, harris2023PRR}. If $w_{\mu\nu} = 0$, we have block-Gaussian matrices, whose spectrum falls inside a spectral disk with radius $\lambda_\text{sr}  = \lambda_\sigma\sqrt{ N / (2M) }$, where $\lambda_\sigma^2$ is the leading eigenvalue of $\mat{\sigma}^2$ \cite{ahmadian_properties_2015, livan_introduction_2018, potters_first_2020}. If $w_{\mu\nu}\neq0$, the bulk is subject to a low-rank perturbation, potentially leading to the emergence of an outlier $\lambda_\text{ot} = N\lambda_w/(2M)$, with $\lambda_w$ the largest eigenvalue of $\mat{w}$ \cite{potters_first_2020, mastrogiuseppe_stochastic_2025}. Therefore, the largest eigenvalue of $\unw$ is related to the reduced matrices as $\lambda_{\tilde W} = \max(\lambda_\text{ot}, \lambda_\text{sr})$.  When the largest eigenvalue is the outlier $\lambda_\text{ot}$, we will say the network is in the \emph{low-rank-dominated phase}, while if the largest eigenvalue is $\lambda_\text{sr}$, we will refer to a \emph{bulk-dominated phase}. Notice that the finite-size scaling exponent depends on the leading eigenvalue, as $\lambda_\text{ot} \sim \mathcal O(N)$ but $\lambda_\text{sr} \sim \mathcal O (\sqrt N)$.  

\begin{figure*}
  \includegraphics[width=\linewidth]{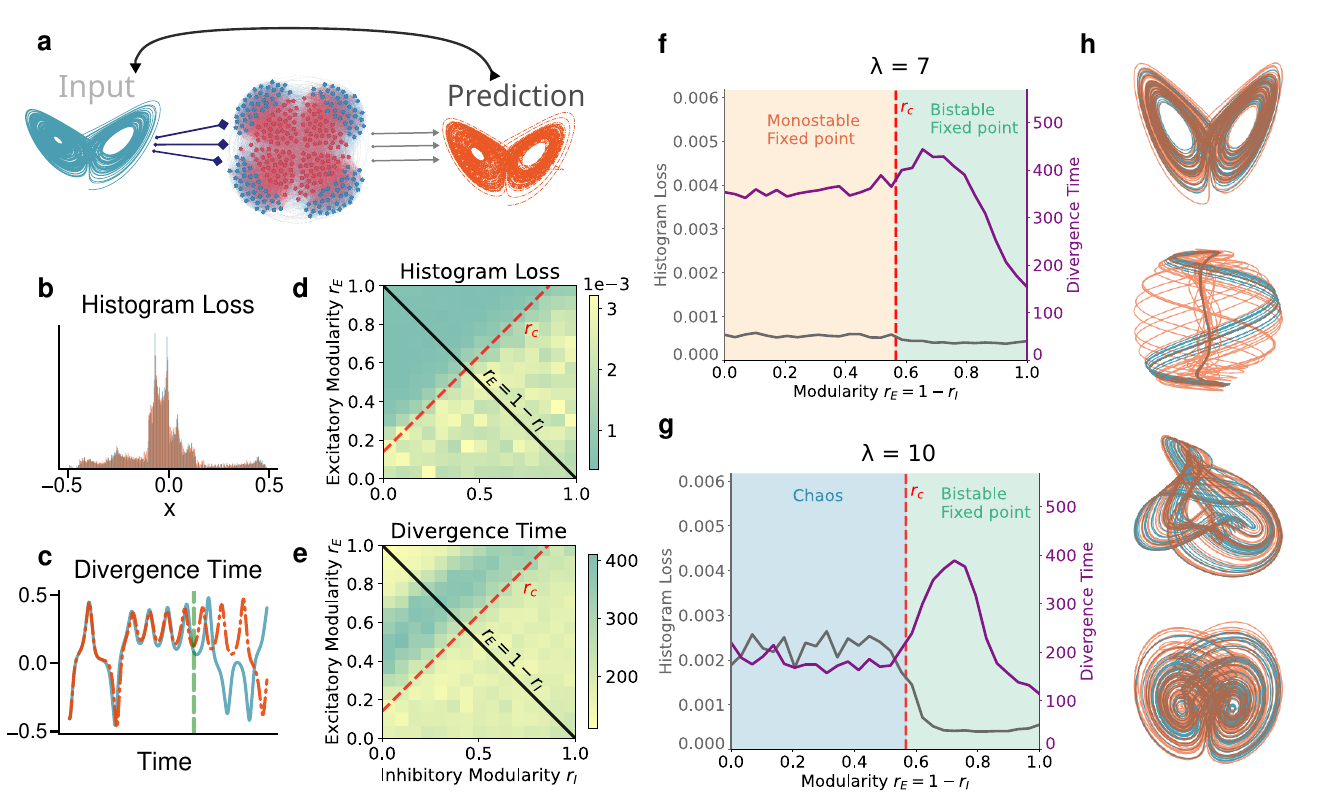}
  \caption{\textbf{The dynamical state impacts the computational performance of the ESN.}\textbf{a.} Schematic of the echo state network (ESN) architecture. \textbf{b.} The histogram loss evaluates the ESN's performance by comparing the distribution of values in the predicted (orange) and target (blue) time series.  \textbf{c.} The divergence time of the predicted and target time series gives us a second measure of the ESN's performance. The plot shows the actual (blue) and predicted (orange) trajectories from the point at which the external input is switched off and the reservoir begins operating as an echo-state network. \textbf{d.} Median (of 500 runs) histogram and \textbf{e.} divergence time losses for different $E$ and $I$ modularity values and a spectral radius of $\lambda = 10$ indicate that, as expected from the analytics, the computational capabilities are optimized near the point of the chaotic transition (Eq. \ref{eq:rec}, marked by the dashed orange line). \textbf{f.} The median (500 runs) histogram and divergence time losses for different levels of $E$ and $I$ modularity across the $r_E = 1 - r_I$ line for $\lambda = 7$. \textbf{g.} Same for $\lambda = 10$. The modularity at which performance changes coincides with the analytically predicted transition by Eq.~(\ref{eq:rc}). Local maximum performance is achieved close to the critical BBP transition controlled by the network topology. We plot here the ESN's performance in reconstructing the Lorenz attractor. \textbf{h.} Reconstructions of four different chaotic attractors (Lorenz, Aizawa, Halvorsen, Chen) by the reservoir. The blue lines show the input data, and the orange lines show the ESN's predictions.
  }
     \label{fig:single_attractor}
\end{figure*}

Whether the network is low-rank dominated or bulk-dominated depends exclusively on the topology of the modules, which is controlled solely by $r_E, r_I$. For $M=2$, the reduced matrices $\mat{w}, \mat{\sigma}$ take a simple form and can be diagonalized exactly. For a sparse network with probability of connection $p$, where each excitatory weight is approximated by a Gaussian with mean $w_0$ and variance $\sigma_w^2$ (see \supp II), the outlier and spectral radius of $\unw$ are given by
\begin{subequations}
\begin{align}
    \lambda_\text{ot} =& \frac{1}{4}p (r_E-r_I)w_0, \\
    \lambda_\text{sr} =&  \sqrt{ \frac{p}{8N}\left[  4p\sigma_w^2+pw_0^2 \left(4- p(2+r_E^2+r_I^2)   \right)   \right]}.
\end{align}
\end{subequations}
The first surprising feature is that, for $r_E > r_I$, we can observe a large, positive, discrete spectral outlier. Therefore, $r_E >r_I$ constitutes a low-rank-dominated phase. This spectrum is typical of an excitatory-dominated network. However, our networks are balanced, with excitation and inhibition having equal weights on average. Thus, even in a balanced network, the topological structure can generate a spectral outlier, making our system behave similarly to an excitation-dominated network. As discussed above, the network cannot undergo a transition to a chaotic regime when such an isolated eigenvalue dominates.
For a chaotic transition to occur, the spectral bulk must first cross the instability threshold, i.e., chaos requires the bulk-dominated phase. This is possible only when the outlier has been merged inside the spectral disk. The first point at which this can happen is when the outlier enters the spectral disk, i.e., at $\lambda_\text{ot} = \lambda_\text{sr}$. For a fixed $r_I$, assuming that $N\gg 1$, this condition leads to a linear relation between $r_E$ and $r_I$,
\begin{equation}
    r_E ^c \simeq r_I + \frac{2\sqrt{ 2\sigma_w^2+(2-p)w_0^2 } }{w_0\sqrt{pN}},
    \label{eq:rec}
\end{equation}
so the bulk-dominated phase appears only for $r_E <r_E ^c$. To further simplify the upcoming dynamical mean-field computations, we restricted ourselves to $r_{E}  \equiv r$ and $r_{I} \equiv 1-r$, so the critical point reads  
\begin{equation}
    r_c \simeq \frac{1}{2} + \frac{\sqrt{2\sigma_w^2 + (2-p)w_0^2}}{w_0\sqrt{Np}}. 
        \label{eq:rc}
\end{equation}

Therefore, strong clustering of excitatory neurons is accompanied by weakly structured inhibition, and vice versa. This setting preserves the overall level of modularity while selectively varying which cell type exhibits more modular connectivity, allowing us to use $r$ as a control parameter while retaining all the dynamical regimes identified in the full parameter space. In \supp~IV, we numerically test that using $r$ and $\lambda$ as control parameters, we capture all the dynamical regimes present in the system. For $r > r_c$, the system is in the low-rank-dominated phase, which becomes unstable when its single outlier crosses the instability threshold. Chaos requires the bulk-dominated phase, i.e., $r <r_c$.  We remark that $r_c$ holds independently of the maximum eigenvalue $\lambda$ of the normalized network. Thus, the transition corresponds to a horizontal line in\fig\ref{fig:dynamics}b.

In random matrix theory, the critical transition at $r_c$ is called a Baik-Ben Arous-Peché (BBP) transition \cite{baik_phase_2005, potters_first_2020}. The finite-size scaling of the system at each side of the transition is different. In the low-rank-dominated phase, the connectivity is normalized by $\lambda_{\tilde W}\sim \mathcal O(N)$, and the fluctuations of $\mat{W}$ vanish in the thermodynamic limit. Therefore, a standard stability analysis suffices, leading to a transition between a monostable fixed point and a bistable one happening at $\lambda_c ^{lr} = 1/\phi'(0) = 4$, with $\phi(x)$ the system nonlinearity (see Methods, \fig\ref{fig:dynamics}b).  On the other hand, in the bulk-dominated regime $\lambda_{\tilde W} \sim \mathcal O (\sqrt N)$. In this case, balance must cancel the low-rank contribution. Otherwise, the input to a neural population will diverge as $\mathcal O (\sqrt N)$ in the thermodynamic limit. Thus, the dynamics depend only on the heterogeneous bulk, which needs to be tackled via DMFT. 

The DMFT approach substitutes all variables by a set of $2M$ representative neurons, one for each population, whose input is described by a random variable. The stability of the monostable fixed point depends on the matrix
\begin{equation}
    S_{\mu \nu} = a_\mu \sigma_{\mu \nu} ^2 \langle \left[  \phi'(y^* _\nu) \right]^2 \rangle,
\end{equation}
where $a_\mu$ is the fraction of neurons in a module $\mu$ (in our case, all $a_\mu = N_\mu/(2M)$),$y^*$ indicates the input current at the fixed point and $\langle \left[  \phi'(y^* _\nu) \right]^2 \rangle$ is an average over the distribution of fixed points. In the DMFT setting, we assume that all nodes in the same module behave identically. Stability is lost when $\lambda^2 \mat{S} /\lambda_\sigma^2 > 1$,  consistently with previous studies \cite{aljadeff2015PRL, kadmon2015PRX} (see \supp II for a complete derivation).

The fact that the low-rank contribution must be balanced, and the assumption that no module differs statistically from others, allows us to obtain a closed expression for the critical point, $\lambda_c^b \simeq6\sqrt\pi=10.68$. This approximation coincides very well with numerical simulations.

Therefore, the phase diagram in \fig\ref{fig:dynamics}b contains 4 phases. The upper part (above $r_c$) corresponds to the low-ranked-dominated regime. This regime contains two phases separated by $\lambda_c ^{lr}$: a monostable fixed point and a bistable fixed point. The bulk-dominated regime (below $r_c$) also contains two phases, separated by $\lambda_c^b$, a monostable fixed point and a chaotic phase. Example trajectories and spectra for each of these phases can be found in \fig\ref{fig:dynamics}c. It is important to note that although for low $\lambda$ both regimes contain a monostable fixed point, these are different phases. In the low-rank-dominated regime, fluctuations from heterogeneity are suppressed, leading to a low-variance distribution for the fixed points. In contrast, the bulk-dominated stable fixed point has high variance. This becomes clearer when we compute the variance of the fixed-point distribution, shown in \fig\ref{fig:dynamics}d alongside theoretical predictions from DMFT. For low $\lambda$ (top panel), the variance of the monostable fixed point collapses at $r_c$, indicating the BBP phase transition. 

\subsection{Computational capabilities depend on the dynamical state}
\begin{figure*}
  \includegraphics[width=\linewidth]{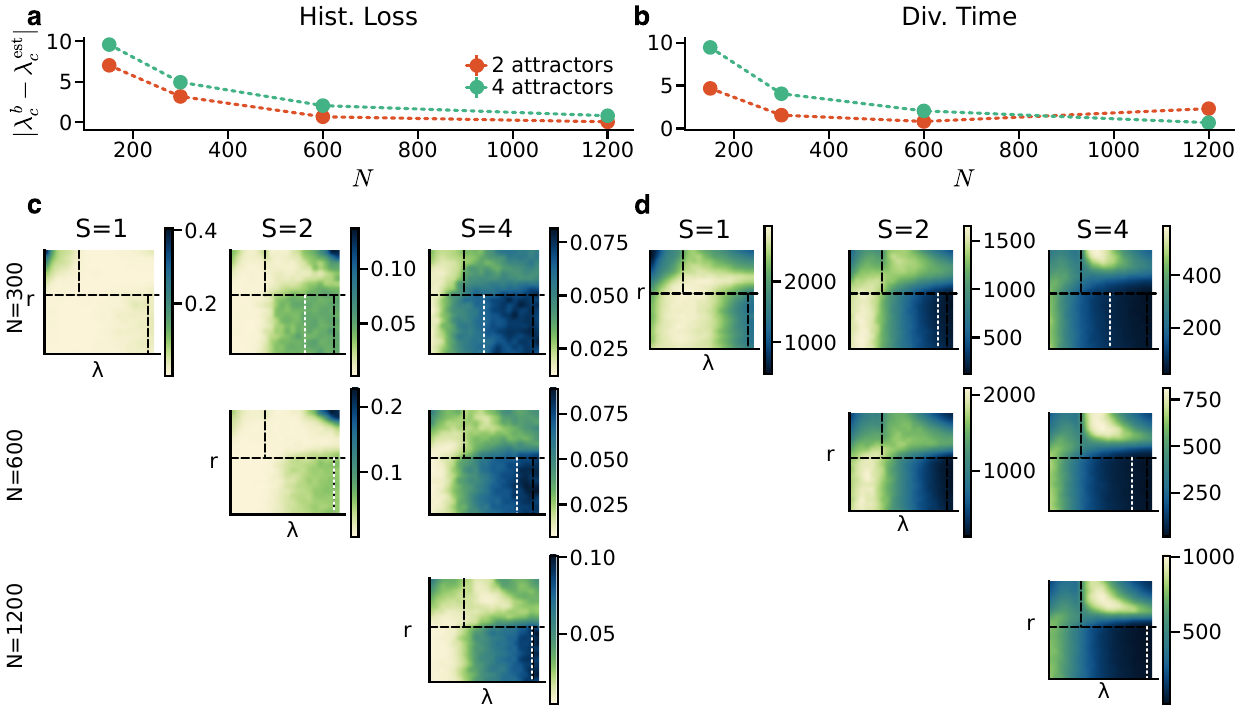}
  \caption{\textbf{The effect of network size and number of predicted attractors.} \textbf{a.} Distance from the estimated transition to the critical spectral radius $\lambda_c ^b$ for increasing system sizes. The network is tasked with reconstructing 2 (red) or 4 (green) chaotic attractors simultaneously. The distance to criticality shifts with the number of attractors, but decreases with the system size. To estimate the transition point, we find the point where the histogram loss increases abruptly for each value of $r$ below $r_c$. The result is averaged over these values of clustering. \textbf{b.} Histogram loss landscape for different system sizes and number of attractors to reconstruct. Theoretical prediction for transitions overlaid as a dashed black line. For 2 and 4 attractors, the estimation of the critical point (corresponds to the dots in panel (a)) is highlighted with a white dotted line.   \textbf{(c),(d)} Same as (a), (b), but using the divergence time instead of the histogram loss. Note that the colormap is inverted, with blue tones indicating generally bad reservoir performance.}
       \label{fig:many_attractors}
\end{figure*}

To assess how modularity influences computational performance, we use the recurrent network as the reservoir in an Echo State Network (ESN) \cite{jaeger2001echo}. In this framework, the recurrent connectivity remains fixed, and only a linear readout is trained. During training, the readout is optimized to predict the next step of a chaotic time series under teacher forcing. During testing, the predicted output is fed back as input, closing the loop and allowing the network to autonomously generate the time series trajectory (\fig\ref{fig:single_attractor}a). We quantify successful learning by the network’s ability to sustain accurate trajectory generation after switching to the autonomous regime.

We evaluate a network's performance using two complementary metrics. First, we compute the distance between the distribution of points in the target and predicted time series (histogram loss \fig\ref{fig:single_attractor}b). Second, we measure the divergence time, defined as the first time point at which the generated trajectory deviates from the target trajectory beyond a prescribed threshold (\fig\ref{fig:single_attractor}c). Because the target dynamics is chaotic, even small prediction errors inevitably lead to trajectory divergence. After divergence, however, the generated trajectory may still remain close to the underlying chaotic attractor. The divergence time, therefore, quantifies how long the network can faithfully track a specific trajectory, whereas the histogram loss assesses how well it captures the geometry of the attractor in phase space. We report both metrics throughout.

First, we evaluate performance on the Lorenz attractor, a three-dimensional chaotic attractor commonly used for time-series prediction. We begin by testing whether performance is affected by different levels of modularity in the $E$ and $I$ recurrent connections, with a fixed spectral radius $\lambda = 10$.

Analytic derivations predict the transition between bulk- and outlier-dominated connectivity at $r_E^c = a + r_I$ with $a\simeq0.6$. Interestingly, the transitions in both performance metrics are situated close to this line (\fig\ref{fig:single_attractor}d,e). We observe that shifting both modularities by the same value (i.e., parallel to the critical line) does not change either histogram loss or divergence time (\fig\ref{fig:single_attractor}d, e). This allowed us to constrain the investigation to the $r_E = 1 - r_I$ line instead of analyzing the full $(r_E, r_I)$ space. Here, the transition to loss of performance happens at a single $r_c$, which is well predicted by Eq.~(\ref{eq:rc}).

Therefore, we study the effect of the E/I clustering on performance along the line $r_E = 1 - r_I = r$ without loss of generality. The resulting loss landscape closely mirrors the dynamical regimes predicted by the DMFT analysis (\fig\ref{fig:single_attractor} f,g) and reveals how network dynamics influence the computational capabilities of the network. In particular, the large-variance monostable fixed point displays generally high performance across the whole phase (\fig\ref{fig:single_attractor}g), whereas the fully chaotic regime is detrimental to performance (\fig\ref{fig:single_attractor}g). For each $\lambda$, the performance along the cut reaches a local maximum at the bistable phase, close to the BBP transition at $r_c$. This effect is most pronounced for the divergence time metric. Remarkably, the BBP transition separates different dynamical regimes at $r_c$, but performance peaks after $r_c$ in both cases, highlighting the role of structure on the network's information-processing capabilities (\fig\ref{fig:single_attractor}f,g). Finally, the global maximum performance happens near the interface between the monostable-bistable transition and not next to the edge of chaos, in contrast with unstructured recurrent networks.

Thus, modularity can tune the network with a sufficiently large spectral radius to a dynamical regime that supports good computational performance. The optimal configurations, characterized by clustered excitation and relatively uniform inhibition, bring the network near the heterogeneous monostable-bistable transition. 
This finding is consistent with established results on reservoir computing, which show that performance is maximized at criticality. Although we see a local maximum in performance at the so-called edge of chaos, as in previous studies, the optimal performance is not found at this point. Moreover, heterogeneity at the monostable point can support highly performant networks. 

\cite{jaeger2001echo, Jaeger2004Science}. 

Finally, to ensure that our results generalize to other chaotic time series and are not specific to the Lorenz attractor  \cite{Lorenz1963}, we repeat our experiment with three other chaotic attractors (Aizawa, Chen, and Halvorsen) \cite{Chen1999, Halvorsen, Aizawa1982GlobalAO}. Despite small changes in the average loss across attractors and small shifts in the loss distributions, the overall patterns remain the same, suggesting that our estimate of the network's dynamical state can effectively predict the region where the ESN's performance is maximized.

\subsection{The simultaneous prediction of multiple chaotic attractors shifts the optimal connectivity structures in small networks}

We test whether the reservoir can \emph{simultaneously} reconstruct multiple independent chaotic time series. Specifically, we present the network with combinations of the four chaotic attractors (Lorenz, Aizawa, Halvorsen, and Chen) \cite{Lorenz1963, Chen1999, Halvorsen, Aizawa1982GlobalAO} we tested individually, creating higher-dimensional inputs (6, 9, or 12-dimensional for 2, 3, or 4 attractors, respectively). While the network generally succeeds in reconstructing time series consisting of combinations of multiple attractors, as would be expected due to the higher complexity of the task, the average performance decreases as the number of simultaneously predicted attractors increases (\fig\ref{fig:many_attractors}b, d). 

We test the network performance for different combinations of $E/I$ modularity (varying $r = r_E = 1 - r_I$) and spectral radius $\lambda$. As we increase the number of presented attractors, the loss map shifts away from the predicted bifurcation diagram, with the chaotic transition appearing at a lower spectral radius $\lambda$. The main contribution to the shift comes from finite-size effects. To better visualize this, we performed experiments with reservoirs of increasing size. Smaller networks not only perform worse but also show a shifted transition to criticality, similar to the shift we observe for fixed-size networks with multiple attractors. We verify that performance increases in the monostable phase as the network size is gradually increased (\fig\ref{fig:many_attractors}a, c). In particular, we find that as we increase network size, the loss values transition of reservoirs predicting multiple attractors (\fig\ref{fig:many_attractors}b, d) slowly moves towards the recurrent network's phase diagram (\fig\ref{fig:dynamics}b). The finite-size shift of a critical point is a well-known effect in the theory of phase transitions \cite{binney_theory_2001}, but here we also observe it as a function of the input dimensionality. It is important to note, however, that for a microscopic network, the sampled low-rank input is not exactly balanced, leading to the appearance of spectral outliers outside the expected radius even in the infinite-size limit (see \supp II). These are not only detrimental to reservoir performance but can also lead to shifts relative to theory predictions, which so far neglect such outliers. 

The loss landscape in \fig\ref{fig:many_attractors} shows that criticality enhances computation capabilities in our system, as losses are reduced close to the bifurcations separating the monostable, bistable, and chaotic regimes. At the interface between the chaotic and bistable regions, the BBP transition, performance peaks locally just above $r_c$. The BBP separation between the two monostable regimes shows a more moderate effect. The monostable-bistable transition also increases performance at the edge of instability. The heterogeneous monostable region (lower left region, $r < r_c$, $\lambda < \lambda_c$) reduces the histogram loss, while the divergence time improves as we move closer to the edge of chaos. Rather than a pronounced peak in performance at the edge of chaos, we have a broader region appearing before the transition. In fact, the best global performance is attained in the bistable regime, close to the corner of the bifurcation point between the four states, i.e., simultaneously close to $r_c$ and $\lambda_c$.

Finally, larger networks are better at time-series prediction. Therefore, by comparing smaller networks, i.e., networks closer to failure, we can more effectively evaluate differences in computing capacity between networks with different dynamics. For example, the bistable region gives the reservoir the ability to predict the time series for the longest time while performing well on the 4-attractor reconstruction. This is clearer at $N=300$ (Fig.~\ref{fig:many_attractors}). Observe that this region emerges for clustered excitation.  Hence, modularity can enhance the reservoir's capabilities relative to its random counterpart. Moreover, the effect seems to extend to very large spectral radii as the network size grows, allowing reservoirs to optimally solve the task with $\lambda \gg \lambda_c$.

\subsection{Performance of the network is mostly independent of the modularity of input connections}

So far, each neuron received randomly selected input from one of the multiple input attractors. We hypothesize that random input may be suboptimal and ask whether structured input targeted to specific $E$ and $I$ populations affects the network's performance.

\begin{figure*}
  \includegraphics[width=\textwidth]{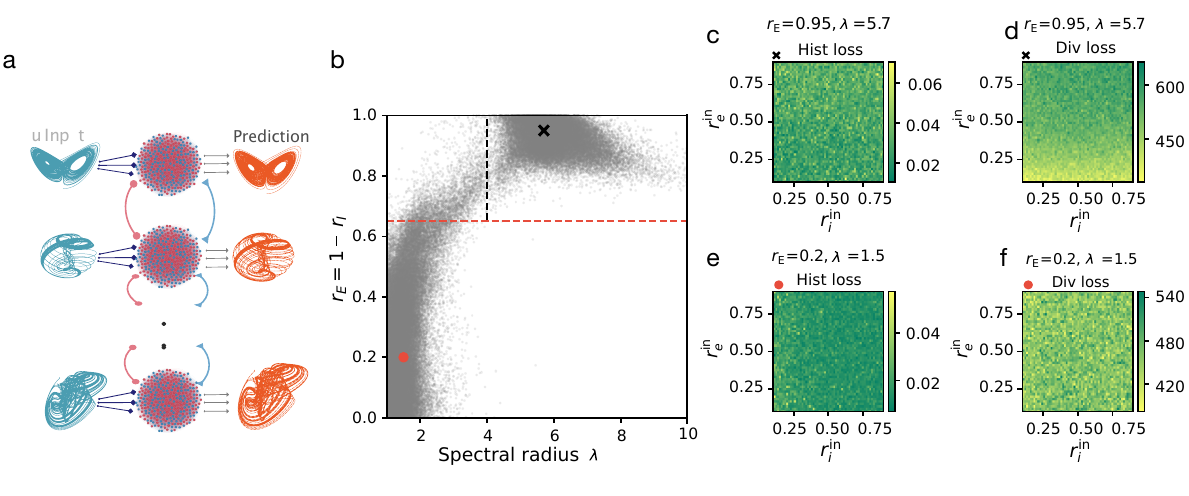}
  \caption{\textbf{Module-specific input connections have a minor effect on the ESN performance.} \textbf{a.} An outline of the network with multiple attractors. We study how to optimally distribute feedforward $E$ and $I$ connections from different attractors to different network modules.  \textbf{b.} The posterior distribution of the network parameters obtained in simulation-based inference. \textbf{c-f.} The reservoir losses for different levels of $E$ and $I$ modularity in the two indicated points of the posterior, circle $r_{\text{E}}=$0.2, $\lambda=5.7$, cross $r_{\text{E}}=$0.95$, \lambda=1.5$,}
 \label{fig:input_sbi}
\end{figure*}
To test the role of input modularity, we simulated a network of four modules that simultaneously reconstruct four chaotic attractors. Each module was assigned one attractor, and input specificity was controlled by excitatory and inhibitory input modularity, $r^{In}_E$ and $r^{In}_I$, where $r^{\text{Input}}=1$ sets fully module-specific input and $r^{\text{Input}}=0$ sets random assignment.

To systematically map the space of optimal solutions, we use Bayesian inference. The added input modularity parameters create many network configurations that can solve the multi-attractor reconstruction task. Because no analytical treatment is available for input-driven networks, we use simulation-based inference (SBI) \cite{goncalves_training_2020,boelts_sbi_2025}. We run $3\times10^5$ simulations for networks with $S=4$ inputs and $N=300$ recurrent units. Then we compute the resulting average histogram and divergence losses across all four attractors as $
\sum_s\text{Hst}_{s} + \frac{1}{1 + \text{Div Time}_{s}},
\qquad s \in \{1,2,3,4\}$ to infer the posterior distributions over spectral radius, recurrent modularity, and input modularity.

The posterior over recurrent parameters (\fig\ref{fig:input_sbi}b) revealed two optimal solutions, consistent with \fig\ref{fig:many_attractors}: one in the bistable regime with high spectral radius and excitatory modularity, and one in the analytically predicted monostable regime with low spectral radius and modularity. 

Next, we investigate the effect of input modularity within the two optimal recurrent architectures (\fig\ref{fig:input_sbi}b and \supp Fig. 3). We stimulate the networks for different excitatory and inhibitory input modularity, while fixing the spectral radius and recurrent modularity to two of the most probable solutions (\fig\ref{fig:input_sbi}c,d,e,f). Input modularity had only a modest effect. In the bistable case (\fig\ref{fig:input_sbi}e,f), excitatory modularity showed a clear positive effect. In the monostable case, inhibitory modularity provided only a minor improvement in histogram loss (\fig\ref{fig:input_sbi}e,f).

\section{Discussion}


In this work, we study how cell-type–specific modular connectivity in balanced E/I networks shapes population dynamics and affects computational capabilities. Using dynamic mean-field theory, we show that varying levels of excitatory and inhibitory modularity control transitions between dynamical regimes, including the transition to chaos. Using networks with different topologies as reservoir computers, we show how these dynamical regimes directly influence performance on a complex temporal prediction task. In particular, networks with strongly modular excitatory connectivity and relatively homogeneous inhibition achieve optimal computational performance near transitions between dynamical regimes.

A central idea linking the dynamical state of the network and computational performance is that maximal processing capacity is achieved when a system operates at the edge of instability \cite{langton1990computation}.  For echo-state reservoir computers, optimal performance has often been linked to the critical point separating stable and chaotic dynamics \cite{legenstein_edge_2007, lukosevicius_reservoir_2009}, although criticality is not a necessary condition for optimal performance \cite{lukosevicius_reservoir_2009, te_vrugt_introduction_2026}. In biological systems, the criticality hypothesis has also been invoked as a simple framework to explain information processing capabilities in the brain \cite{munoz_colloquium:_2018}, including positive experimental support \cite{beggs_neuronal_2003, friedman_universal_2012, fontenele_criticality_2019}. Moreover, some results suggest that the optimal state should be subcritical, leaving room for an external drive without destabilizing the system \cite{priesemann_spike_2014, wilting_between_2019}, similar to the findings in echo-state networks.  

Our results suggest that criticality is favorable for computation in the thermodynamic limit. We find that computational capabilities are enhanced close to all phase transitions in the model, although in different ways. Notably, the interface between the large-variance monostable point and the chaotic state, commonly known as the 'edge-of-chaos', does not have a pronounced maximum of performance before the critical point, as in the unstructured rate network case \cite{lukosevicius_reservoir_2009}. Instead, it appears smeared over a larger region. In contrast, we find performance peaks along the BBP transition happening at $r_c$, highlighting the important role of excitation-inhibition clustering for information processing. In particular, optimal performance occurs in the bistable state, near both the BBP $r_c$ and the edge-of-instability $\lambda_c ^{lr}$. At this point, the edge of chaos is not yet present. 

Classical reservoir computing relies on uniformly connected recurrent networks \cite{jaeger2001echo}, which, while analytically convenient, do not reflect the structured organization observed in biological circuits. Cortical networks exhibit pronounced modular and clustered connectivity, with distinct patterns across excitatory and inhibitory populations, and such structure has been shown to give rise to distinct dynamical regimes \cite{LitwinKumar2012natneuro} and influence learning \cite{Giannakakis2023}.
Our results build on this by demonstrating that introducing cell-type–specific modularity not only preserves computational functionality but also provides an additional degree of control over network dynamics. In particular, the regime we identify as computationally optimal, characterized by strongly modular excitatory connectivity and relatively homogeneous inhibition, is consistent with experimentally observed motifs in cortical circuits, where excitatory neurons tend to form clustered, feature-specific assemblies, while inhibitory connectivity is broader and less selective \citep{scholl2019functional, znamenskiy2024functional}. This suggests that topological asymmetry between excitation and inhibition may help drive biological networks toward dynamical regimes favorable for information processing.


The application of DMFT to structured systems with excitatory and inhibitory populations is an active area of research. Our approach is based on theoretical developments for multipopulation neural networks \cite{aljadeff2015PRL, kadmon2015PRX, mastrogiuseppe2017PCB}. However, none of these studies explicitly examine the effect of modular connectivity. Our approach follows Kadmon and Somponlinsky by approximating the modular connectivity matrix with Gaussian blocks \cite{kadmon2015PRX} and applying the theory to elucidate the effect of topology. Other works in this line, such as Mastrogiuseppe and Ostojic \cite{mastrogiuseppe2017PCB}, study E/I dynamics in detail, while Aljadeff \emph{et al.} \cite{aljadeff2015PRL} consider only Gaussian networks with vanishing mean, and thus no excitatory or inhibitory behavior. A large family of structured networks was studied by Ahmadian \emph{et al.} \cite{ahmadian_properties_2015}, but our model does not fall into this class of problems. Our results could be useful for other scenarios. Finally, Clark and Beiran \cite{clark_structure_2025} recently studied a multi-region model in which communication among modules occurs via low-rank, structured connectivity. However, they do not enforce Dale's law or positive firing rates, limiting the biological plausibility of their results. Hence, our results complement and extend the previous literature. 

Moreover, we see that the balance state plays a fundamental role, even in the presence of structured networks. This issue arises because the low-rank contribution coming from structured connectivity does not scale the same way as the bulk \cite{potters_first_2020}. To address this problem, many DMFT studies normalize the low-rank and bulk components separately \cite{kadmon2015PRX, mastrogiuseppe2018N, schuessler_dynamics_2020, clark_structure_2025}. We show that this is not necessary. In particular, in the bulk-dominated regime, the input current from the low-rank contribution scales with system size, apparently leading to divergence. However, we show that network balance is able to cancel such divergences, as found in the classical literature on balanced states \cite{vanVreeswijk1996, haim_sompolinsky_chaos_1988}.  

The BBP transition may be relevant to the study of brain dynamics, bringing together the two types of criticality discussed by Dahmen \cite{dahmen2019PNAS}. A vast literature already highlights the importance of the balanced state, which appears in the bulk-dominated regime. More recently, the study of low-rank recurrent neural networks has attracted a lot of attention \cite{mastrogiuseppe2018N, mastrogiuseppe_stochastic_2025, clark_structure_2025}.  A modular network near a BBP transition, equipped with a plasticity mechanism, could switch between low-rank-dominated and bulk-dominated behavior, exploiting advantages from both states. In fact, our reservoirs seem to be generally efficient close to the BBP transition. Whether such a mechanism exists in the brain, and its relevance, remains an open question.


The theory also presents several important limitations. First, we keep reservoirs small to ensure reasonable training times. This leaves each module with only a handful of neurons. However, the theory is studied in the thermodynamic limit, for infinite sizes. Moreover, we compare the network's autonomous behavior with the input-driven reservoir. Both effects contribute to shifting the location of the phase transitions relative to theoretical predictions. To fully describe this phenomenon, one should characterize the dynamics of the finite-size recurrent network under chaotic input. This remains an open theoretical question that falls outside the scope of our work and should be addressed separately. 

A natural next step is to apply this framework to anatomical connectomes extracted from experimental data, using reservoir computing as a probe to relate measured connectivity motifs to dynamical regimes and computational capacity. More broadly, our results suggest that biologically plausible RNNs are a fruitful, and still underused, lens through which computational neuroscience can study the interplay between structure, dynamics, and function.

\section{Methods}

\subsection{Recurrent network}

The dynamics of the network's activity $x = (x_1, \dots, x_N)$ are given by the following equations
\begin{equation}
        \dot x_i = -x_i + \phi \left( \sum_{j=1} ^N W_{ij} x_j + s_i(t)  \right),
\end{equation}
where $\mat{W}$ is the recurrent connectivity matrix and $s_i(t)$ is the external input to each node.
The nonlinearity $\phi$ is the sigmoid
\begin{equation}
\phi(x) = \frac{1}{1 + e^{-x}}.
\end{equation}
Note that in our setup, $\phi(0) = 1/2 \equiv \phi_0$. Initial conditions for each $x_i$ are drawn randomly in the interval $[0,1]$. We employ the Euler algorithm for integration with $\Delta t=0.01$. We have checked that the results are numerically stable for the selected time step.

\subsection{Connectivity}

The matrix $\mat{W}$ contains $M$ modules. Each module consists of an excitatory and inhibitory population, making a total of $2M$ populations. The network is sparse, and the connection probability from a neuron $i$ in population $\mu$ to a neuron $j$ in population $\nu$ is given by Eq.~(\ref{eq:probconn}). 

We draw connection weights from a uniform random distribution on $[w_0, w_0+\Delta w]$. In simulations, we choose $w_0=\Delta w=0.2$. This is common to all populations, but inhibitory connections are multiplied by $-1$. After constructing the matrix $\mat{W}$, we normalize it to have the desired largest eigenvalue $\lambda$.

For theoretical calculations, we approximate the sparse, weighted networks by a Gaussian distribution. To do so, the mean and variance of the uniform distribution $w_0$, $\sigma_w^2 = (\Delta w)^2/12$ are employed. Please refer to \supp II for a detailed derivation.  

\subsection{Input}
Each neuron receives as input one dimension of a $D$-dimensional time series $\vec C(t)$:
\begin{equation}
\vec s(t) = \mat{W}_{\text{Input}} \vec C(t),
\end{equation}
where $\mat{W}_{\text{Input}}$ is an $N\times D$ matrix, with a single non-zero entry per row with weight $w_\text{Input} = 0.5$. Thus, each node receives a single dimension from the time series.

\subsection{Echo state network (ESN)}

The simulation of the ESN is divided in two phases:

\hspace{0.65cm}

\textbf{Stimulus-driven:} The time series $\vec C(t)$ is given as input for $T$ time steps, leading to the recurrent network activity $\mat{X} = [\vec x_1, \dots, \vec x_T ]$ obtained by stacking network's state at each time. Using Ridge regression, a linear layer $\mat{W}_R$ is trained to predict the trajectory of the time series. Thus, using the network activity at a given time step, $\vec x_t$ the linear layer predicts the next step of the time series $\vec C(t+1) \equiv \vec C_{t+1}$. The target is constructed by stacking the time series vectors in columns, $\mat{y}^{\text{target}} = [\vec C_2, \dots, \vec C_{T+1} ]$, and regression yields 
\begin{equation}
    \mat{W}_R = \mat{y}^{\text{target}}  \mat{X}^\dagger (\mat{X} \mat{X}^\dagger + \beta \mat{I})^{-1},
\end{equation}
where $\beta$ is the Ridge regression's normalization constant and $\dagger$ stands for transpose.

\hspace{1cm}

\textbf{Closed-Loop:} Once the training is complete, we replace the input from the given time series $\vec C_{T+1}$, with the prediction generated via the trained linear layer $\mat{W}_R x^\dagger _T)$. If Ridge regression was successful, then $\vec x'_{T+1} = W_R x^\dagger _T$ and the recurrent network will produce the same activity as if its input had been the actual time series. 

We continue in the same way for $T'$ time steps, using the linear layer's prediction as the next input for the recurrent network, creating a fully closed loop which results in the reconstructed time series $\mat{X'} = [ \vec x'_{T+1}, \ldots, \vec x' _{T+T'}]$ 

\subsection{Evaluation}

To evaluate the performance of the ESN, we compare the reconstructed time series $\mat{X'}$ with the actual corresponding time series $\mat{Y} = \{C_{T+1}, \dots, C_{T+T'} \} $. In an ideally functioning ESN, the dynamics of the reservoir should be rich enough to capture the trajectory of the time series and enable a very precise reconstruction, so that $\mat{X'} \approx \mat{Y}$ over the entire testing period. We use two metrics to quantify the ESN performance:

\textbf{Divergence Time.} We estimate the time $\tau$ at which the reconstructed trajectory and the original one start to diverge, by obtaining the $\tau$ such that $|C_\tau - x'_\tau| > \epsilon$ for a given small threshold $\epsilon$. 
 
\textbf{Histogram Loss.} We compute the histogram of each dimension of the time series separately, both for the original series and the reconstruction. Then, we compute the mean-square difference between the histograms for each dimension and sum the results, multiplied by 100 to avoid vanishing values. 

Each metric measures a different aspect of reservoir prediction. On one hand, divergence time captures the quality of the prediction for individual time series. A long divergence time means the reservoir can imitate the behavior of the chaotic system for a long time. However, it does not provide information about the statistical properties of the reconstruction after divergence. On the other hand, histogram loss can be very small as long as the reservoir trajectory is predicting the stable basin of the chaotic attractor. Notice that one can have a short divergence time with small histogram loss, meaning a reservoir might correctly predict the chaotic basin of attraction without performing well on individual series, but both metrics are usually correlated.

\subsection{Simulation-Based Inference}
We use simulation-based inference \cite{goncalves_training_2020} to find the optimal excitatory and inhibitory input modularity ($r^{in}_E$ and $r^{in}_I$) together with the spectral radius ($\lambda$) and recurrent modularity ($r$). We randomly sampled 250,000 parameters from a uniform prior distribution ($r$, $r^{in}_E$, and $r^{in}_I$ between 0 and 1; $\lambda$ between 1 and 10). We then fed four attractors to the reservoir simultaneously and computed their divergence time and histogram loss. We then obtain an average loss for all of them as $L^{\text{total}} =\sum_s\text{Hst}_{s} + \frac{1}{1 + \text{Div Time}_{s}}$. Next, we fit a normalizing spline flow as a neural posterior estimator to predict the distribution of four parameters given the loss function ($p(r^{in}_E,r^{in}_I,r,\lambda | L^{\text{total}})$). Next, we sampled from this posterior given $L=0$. For Fig.5 c,d,e,g, we run a simulations for a $r_e^{in}$, $r_I^{in}$ between 0.1 and 0.9 with selected values of $r_{E}$ and $\lambda$ .

\section{Acknowledgments}

This work was supported by a Sofja Kovalevskaja Award
from the Alexander von Humboldt Foundation. EG thanks the International Max Planck Research School for Intelligent Systems (IMPRS-IS) for support, and OV thanks the International Max Planck Research School for the Mechanisms of Mental Function and Dysfunction and the Joachim Herz Foundation for support. We acknowledge support from the BMBF through the T\"ubingen AI Center (FKZ: 01IS18039B).
AL is a member of the Cluster of Excellence ``Machine Learning: New Perspectives for Science'', EXC 2064/1 (project number 39072764). EG is funded by the Deutsche Forschungsgemeinschaft (DFG, German Research Foundation)  – project number 575506470.
VB acknowledges funding by NextGenerationEU, in the framework of the FAIR - Future Artificial Intelligence Research project (FAIR PE00000013 – CUP B43C22000800006).
We thank Enrico Ventura for useful discussions on random matrix theory and Miguel A. Muñoz for feedback on the manuscript.
\bibliographystyle{unsrt}
\bibliography{bibliography}

@article{boelts_sbi_2025,
	title = {sbi reloaded: a toolkit for simulation-based inference workflows},
	volume = {10},
	issn = {2475-9066},
	url = {https://joss.theoj.org/papers/10.21105/joss.07754},
	doi = {10.21105/joss.07754},
	shorttitle = {sbi reloaded},
	pages = {7754},
	number = {108},
	journaltitle = {Journal of Open Source Software},
	author = {Boelts, Jan and Deistler, Michael and Gloeckler, Manuel and Tejero-Cantero, Álvaro and Lueckmann, Jan-Matthis and Moss, Guy and Steinbach, Peter and Moreau, Thomas and Muratore, Fabio and Linhart, Julia and Durkan, Conor and Vetter, Julius and Miller, Benjamin Kurt and Herold, Maternus and Ziaeemehr, Abolfazl and Pals, Matthijs and Gruner, Theo and Bischoff, Sebastian and Krouglova, Nastya and Gao, Richard and Lappalainen, Janne K. and Mucsányi, Bálint and Pei, Felix and Schulz, Auguste and Stefanidi, Zinovia and Rodrigues, Pedro and Schröder, Cornelius and Zaid, Faried Abu and Beck, Jonas and Kapoor, Jaivardhan and Greenberg, David S. and Gonçalves, Pedro J. and Macke, Jakob H.},
	date = {2025-04-08},
}

@article{goncalves_training_2020,
	title = {Training deep neural density estimators to identify mechanistic models of neural dynamics},
	volume = {9},
	issn = {2050-084X},
	url = {https://doi.org/10.7554/eLife.56261},
	doi = {10.7554/eLife.56261},
	pages = {e56261},
	journaltitle = {{eLife}},
	author = {Gonçalves, Pedro J and Lueckmann, Jan-Matthis and Deistler, Michael and Nonnenmacher, Marcel and Öcal, Kaan and Bassetto, Giacomo and Chintaluri, Chaitanya and Podlaski, William F and Haddad, Sara A and Vogels, Tim P and Greenberg, David S and Macke, Jakob H},
	editor = {Huguenard, John R and O'Leary, Timothy and Goldman, Mark S},
	urldate = {2022-07-21},
	date = {2020-09-17},
	note = {Publisher: {eLife} Sciences Publications, Ltd},
}

@article{hilgetag2004clustered,
  title={Clustered organization of cortical connectivity},
  author={Hilgetag, Claus C and Kaiser, Marcus},
  journal={Neuroinformatics},
  volume={2},
  pages={353--360},
  year={2004},
  publisher={Springer}
}

@article{kirsch2018modular,
  title={Modular networks: Learning to decompose neural computation},
  author={Kirsch, Louis and Kunze, Julius and Barber, David},
  journal={Advances in neural information processing systems},
  volume={31},
  year={2018}
}

@article{houk1994information,
  title={Information processing in modular circuits linking basal ganglia and cerebral cortex},
  author={Houk, James C},
  year={1994}
}

@article{lundqvist2006attractor,
  title={Attractor dynamics in a modular network model of neocortex},
  author={Lundqvist, Mikael and Rehn, Martin and Djurfeldt, Mikael and Lansner, Anders},
  journal={Network: Computation in Neural Systems},
  volume={17},
  number={3},
  pages={253--276},
  year={2006},
  publisher={Taylor \& Francis}
}

@article{yamamoto2018impact,
  title={Impact of modular organization on dynamical richness in cortical networks},
  author={Yamamoto, Hideaki and Moriya, Satoshi and Ide, Katsuya and Hayakawa, Takeshi and Akima, Hisanao and Sato, Shigeo and Kubota, Shigeru and Tanii, Takashi and Niwano, Michio and Teller, Sara and others},
  journal={Science advances},
  volume={4},
  number={11},
  pages={eaau4914},
  year={2018},
  publisher={American Association for the Advancement of Science}
}

@article{pradhan2011modular,
  title={Modular organization enhances the robustness of attractor network dynamics},
  author={Pradhan, Neeraj and Dasgupta, Subinay and Sinha, Sitabhra},
  journal={Europhysics Letters},
  volume={94},
  number={3},
  pages={38004},
  year={2011},
  publisher={IOP Publishing}
}

@article{galstyan2007cascading,
  title={Cascading dynamics in modular networks},
  author={Galstyan, Aram and Cohen, Paul},
  journal={Physical Review E—Statistical, Nonlinear, and Soft Matter Physics},
  volume={75},
  number={3},
  pages={036109},
  year={2007},
  publisher={APS}
}

@article{blanco2024population,
  title={Population bursts in a modular neural network as a mechanism for synchronized activity in KNDy neurons},
  author={Blanco, Wilfredo and Tabak, Joel and Bertram, Richard},
  journal={PLOS Computational Biology},
  volume={20},
  number={7},
  pages={e1011820},
  year={2024},
  publisher={Public Library of Science San Francisco, CA USA}
}

@article{valencia2009complex,
  title={Complex modular structure of large-scale brain networks},
  author={Valencia, Miguel and Pastor, MA and Fern{\'a}ndez-Seara, MA and Artieda, J and Martinerie, J and Chavez, M},
  journal={Chaos: An Interdisciplinary Journal of Nonlinear Science},
  volume={19},
  number={2},
  year={2009},
  publisher={AIP Publishing}
}

@article{sporns2016modular,
  title={Modular brain networks},
  author={Sporns, Olaf and Betzel, Richard F},
  journal={Annual review of psychology},
  volume={67},
  number={1},
  pages={613--640},
  year={2016},
  publisher={Annual Reviews}
}

@article {Giannakakis2023,
	author = {Emmanouil Giannakakis and Oleg Vinogradov and Victor Buend{\'\i}a and Anna Levina},
	title = {The topology of E/I recurrent networks regulates the effects of synaptic plasticity},
	elocation-id = {2023.02.27.530253},
	year = {2023},
	doi = {10.1101/2023.02.27.530253},
	publisher = {Cold Spring Harbor Laboratory},
	URL = {https://www.biorxiv.org/content/early/2023/07/11/2023.02.27.530253},
	eprint = {https://www.biorxiv.org/content/early/2023/07/11/2023.02.27.530253.full.pdf},
	journal = {bioRxiv}
}

@article{jaeger2001echo,
  title={The “echo state” approach to analysing and training recurrent neural networks-with an erratum note},
  author={Jaeger, Herbert},
  journal={Bonn, Germany: German National Research Center for Information Technology GMD Technical Report},
  volume={148},
  number={34},
  pages={13},
  year={2001},
  publisher={Bonn}
}

@article{Jaeger2004Science,
author = {Jaeger, Herbert and Haas, Harald},
year = {2004},
month = {05},
pages = {78-80},
title = {Harnessing Nonlinearity: Predicting Chaotic Systems and Saving Energy in Wireless Communication},
volume = {304},
journal = {Science (New York, N.Y.)},
doi = {10.1126/science.1091277},
}

@article{scholl2019functional,
  title={Functional logic of layer 2/3 inhibitory connectivity in the ferret visual cortex},
  author={Scholl, Benjamin and Wilson, Daniel E and Jaepel, Juliane and Fitzpatrick, David},
  journal={Neuron},
  volume={104},
  number={3},
  pages={451--457},
  year={2019},
  publisher={Elsevier}
}

@inproceedings{caluwaerts20132IJCNNI,
  title = {The Spectral Radius Remains a Valid Indicator of the {{Echo}} State Property for Large Reservoirs},
  booktitle = {The 2013 {{International Joint Conference}} on {{Neural Networks}} ({{IJCNN}})},
  author = {Caluwaerts, Ken and Wyffels, Francis and Dieleman, Sander and Schrauwen, Benjamin},
  year = {2013},
  month = aug,
  pages = {1--6},
  issn = {2161-4407},
  doi = {10.1109/IJCNN.2013.6706899}
}

@article{jiang2019PRR,
  title = {Model-Free Prediction of Spatiotemporal Dynamical Systems with Recurrent Neural Networks: {{Role}} of Network Spectral Radius},
  shorttitle = {Model-Free Prediction of Spatiotemporal Dynamical Systems with Recurrent Neural Networks},
  author = {Jiang, Junjie and Lai, Ying-Cheng},
  year = {2019},
  month = oct,
  journal = {Physical Review Research},
  volume = {1},
  number = {3},
  pages = {033056},
  publisher = {{American Physical Society}},
  doi = {10.1103/PhysRevResearch.1.033056},
  urldate = {2023-09-08}
}

@incollection{lukosevicius2012NNTotTSE,
  title = {A {{Practical Guide}} to {{Applying Echo State Networks}}},
  booktitle = {Neural {{Networks}}: {{Tricks}} of the {{Trade}}: {{Second Edition}}},
  author = {Luko{\v s}evi{\v c}ius, Mantas},
  editor = {Montavon, Gr{\'e}goire and Orr, Genevi{\`e}ve B. and M{\"u}ller, Klaus-Robert},
  year = {2012},
  series = {Lecture {{Notes}} in {{Computer Science}}},
  pages = {659--686},
  publisher = {{Springer}},
  address = {{Berlin, Heidelberg}},
  doi = {10.1007/978-3-642-35289-8_36},
  urldate = {2023-09-08},
  isbn = {978-3-642-35289-8},
  langid = {english}
}

@article{yildiz2012NN,
  title = {Re-Visiting the Echo State Property},
  author = {Yildiz, Izzet B. and Jaeger, Herbert and Kiebel, Stefan J.},
  year = {2012},
  month = nov,
  journal = {Neural Networks},
  volume = {35},
  pages = {1--9},
  issn = {0893-6080},
  doi = {10.1016/j.neunet.2012.07.005},
  urldate = {2023-09-08}
}

@article{dahmen2019PNAS,
  title = {Second Type of Criticality in the Brain Uncovers Rich Multiple-Neuron Dynamics},
  author = {Dahmen, David and Gr{\"u}n, Sonja and Diesmann, Markus and Helias, Moritz},
  year = {2019},
  month = jun,
  journal = {Proceedings of the National Academy of Sciences},
  volume = {116},
  number = {26},
  pages = {13051--13060},
  publisher = {{Proceedings of the National Academy of Sciences}},
  doi = {10.1073/pnas.1818972116},
  urldate = {2023-11-27}
}

@article{aljadeff2015PRL,
  title = {Transition to {{Chaos}} in {{Random Networks}} with {{Cell-Type-Specific Connectivity}}},
  author = {Aljadeff, Johnatan and Stern, Merav and Sharpee, Tatyana},
  year = {2015},
  month = feb,
  journal = {Physical Review Letters},
  volume = {114},
  number = {8},
  pages = {088101},
  publisher = {{American Physical Society}},
  doi = {10.1103/PhysRevLett.114.088101},
  urldate = {2023-10-02}
}

@article{clark_structure_2025,
	title = {Structure of activity in multiregion recurrent neural networks},
	volume = {122},
	url = {https://www.pnas.org/doi/10.1073/pnas.2404039122},
	doi = {10.1073/pnas.2404039122},
	number = {10},
	urldate = {2026-01-28},
	journal = {Proceedings of the National Academy of Sciences},
	publisher = {Proceedings of the National Academy of Sciences},
	author = {Clark, David G. and Beiran, Manuel},
	month = mar,
	year = {2025},
	pages = {e2404039122},
}

@article{mastrogiuseppe_stochastic_2025,
	title = {Stochastic activity in low-rank recurrent neural networks},
	volume = {21},
	issn = {1553-7358},
	url = {https://journals.plos.org/ploscompbiol/article?id=10.1371/journal.pcbi.1013371},
	doi = {10.1371/journal.pcbi.1013371},
	language = {en},
	number = {8},
	urldate = {2026-06-11},
	journal = {PLOS Computational Biology},
	publisher = {Public Library of Science},
	author = {Mastrogiuseppe, Francesca and Carmona, Joana and Machens, Christian K.},
	month = aug,
	year = {2025},
	pages = {e1013371},
}

@article{baik_phase_2005,
	title = {Phase transition of the largest eigenvalue for nonnull complex sample covariance matrices},
	volume = {33},
	issn = {0091-1798, 2168-894X},
	url = {https://projecteuclid.org/journals/annals-of-probability/volume-33/issue-5/Phase-transition-of-the-largest-eigenvalue-for-nonnull-complex-sample/10.1214/009117905000000233.full},
	doi = {10.1214/009117905000000233},
	language = {en},
	number = {5},
	urldate = {2026-06-11},
	journal = {The Annals of Probability},
	publisher = {Institute of Mathematical Statistics},
	author = {Baik, Jinho and Arous, Gérard Ben and Péché, Sandrine},
	month = sep,
	year = {2005},
	pages = {1643--1697},
}

@article{schuessler_dynamics_2020,
	title = {Dynamics of random recurrent networks with correlated low-rank structure},
	volume = {2},
	url = {https://link.aps.org/doi/10.1103/PhysRevResearch.2.013111},
	doi = {10.1103/PhysRevResearch.2.013111},
	number = {1},
	urldate = {2026-06-12},
	journal = {Physical Review Research},
	publisher = {American Physical Society},
	author = {Schuessler, Friedrich and Dubreuil, Alexis and Mastrogiuseppe, Francesca and Ostojic, Srdjan and Barak, Omri},
	month = feb,
	year = {2020},
	pages = {013111},
}

@article{haim_sompolinsky_chaos_1988,
	title = {Chaos in {Random} {Neural} {Networks}},
	volume = {61},
	doi = {10.1103/PhysRevLett.61.259},
	number = {3},
	journal = {Physical Review Letters},
	author = {{Haim Sompolinsky} and {Andrea Crisanti}},
	year = {1988},
	pages = {259--262},
}

@article{munoz_colloquium:_2018,
	title = {Colloquium: {Criticality} and dynamical scaling in living systems},
	volume = {90},
	doi = {10.1103/RevModPhys.90.031001},
	number = {3},
	journal = {Rev. Mod. Phys.},
	publisher = {APS},
	author = {Muñoz, Miguel A.},
	month = jul,
	year = {2018},
	note = {tex.ids: munoz2018RMPa
tex.numpages: 30
tex.publisher: American Physical Society
numPages: 30},
	pages = {031001},
}

@incollection{te_vrugt_introduction_2026,
	address = {Cham},
	title = {An {Introduction} to {Reservoir} {Computing}},
	isbn = {978-3-032-04129-6},
	url = {https://doi.org/10.1007/978-3-032-04129-6_4},
	doi = {10.1007/978-3-032-04129-6_4},
	language = {en},
	urldate = {2026-04-21},
	booktitle = {Artificial {Intelligence} and {Intelligent} {Matter}: {Nanoscience}, {Soft} {Matter}, {Philosophy}},
	publisher = {Springer Nature Switzerland},
	author = {te Vrugt, Michael},
	editor = {te Vrugt, Michael},
	year = {2026},
	pages = {53--66},
}

@article{lukosevicius_reservoir_2009,
	title = {Reservoir computing approaches to recurrent neural network training},
	volume = {3},
	issn = {1574-0137},
	url = {https://www.sciencedirect.com/science/article/pii/S1574013709000173},
	doi = {10.1016/j.cosrev.2009.03.005},
	number = {3},
	urldate = {2026-04-21},
	journal = {Computer Science Review},
	author = {Lukoševičius, Mantas and Jaeger, Herbert},
	month = aug,
	year = {2009},
	pages = {127--149},
}

@article{legenstein_edge_2007,
	title = {Edge of chaos and prediction of computational performance for neural circuit models},
	volume = {20},
	issn = {0893-6080},
	doi = {10.1016/j.neunet.2007.04.017},
	language = {eng},
	number = {3},
	journal = {Neural Networks: The Official Journal of the International Neural Network Society},
	author = {Legenstein, Robert and Maass, Wolfgang},
	month = apr,
	year = {2007},
	pages = {323--334},
}

@article{fontenele_criticality_2019,
	title = {Criticality between cortical states},
	volume = {122},
	number = {20},
	journal = {Physical review letters},
	author = {Fontenele, Antonio J and de Vasconcelos, Nivaldo AP and Feliciano, Thaís and Aguiar, Leandro AA and Soares-Cunha, Carina and Coimbra, Bárbara and Dalla Porta, Leonardo and Ribeiro, Sidarta and Rodrigues, Ana João and Sousa, Nuno and {others}},
	year = {2019},
	note = {tex.publisher: APS},
	pages = {208101},
}

@article{beggs_neuronal_2003,
	title = {Neuronal avalanches in neocortical circuits},
	volume = {23},
	number = {35},
	journal = {Journal of neuroscience},
	publisher = {Soc Neuroscience},
	author = {Beggs, John M and Plenz, Dietmar},
	year = {2003},
	note = {tex.ids: beggs2003JNa
tex.publisher: Soc Neuroscience},
	pages = {11167--11177},
}

@article{friedman_universal_2012,
	title = {Universal critical dynamics in high resolution neuronal avalanche data},
	volume = {108},
	number = {20},
	journal = {Physical review letters},
	author = {Friedman, Nir and Ito, Shinya and Brinkman, Braden AW and Shimono, Masanori and DeVille, RE Lee and Dahmen, Karin A and Beggs, John M and Butler, Thomas C},
	year = {2012},
	note = {tex.publisher: APS},
	pages = {208102},
}

@article{wilting_between_2019,
	title = {Between {Perfectly} {Critical} and {Fully} {Irregular}: {A} {Reverberating} {Model} {Captures} and {Predicts} {Cortical} {Spike} {Propagation}},
	volume = {29},
	issn = {1047-3211},
	shorttitle = {Between {Perfectly} {Critical} and {Fully} {Irregular}},
	url = {https://academic.oup.com/cercor/article/29/6/2759/5476016},
	doi = {10.1093/cercor/bhz049},
	language = {en},
	number = {6},
	urldate = {2020-11-18},
	journal = {Cerebral Cortex},
	publisher = {Oxford Academic},
	author = {Wilting, J. and Priesemann, V.},
	month = jun,
	year = {2019},
	pages = {2759--2770},
}

@article{priesemann_spike_2014,
	title = {Spike avalanches in vivo suggest a driven, slightly subcritical brain state},
	volume = {8},
	issn = {1662-5137},
	url = {https://www.frontiersin.org/articles/10.3389/fnsys.2014.00108/full},
	doi = {10.3389/fnsys.2014.00108},
	language = {English},
	urldate = {2020-10-27},
	journal = {Frontiers in Systems Neuroscience},
	publisher = {Frontiers},
	author = {Priesemann, Viola and Wibral, Michael and Valderrama, Mario and Pröpper, Robert and Le Van Quyen, Michel and Geisel, Theo and Triesch, Jochen and Nikolić, Danko and Munk, Matthias H. J.},
	year = {2014},
	note = {tex.ids: priesemann2014FSNa},
}

@book{livan_introduction_2018,
	address = {Cham},
	series = {{SpringerBriefs} in {Mathematical} {Physics}},
	title = {Introduction to {Random} {Matrices}},
	volume = {26},
	copyright = {http://www.springer.com/tdm},
	isbn = {978-3-319-70883-6 978-3-319-70885-0},
	url = {http://link.springer.com/10.1007/978-3-319-70885-0},
	urldate = {2025-11-20},
	publisher = {Springer International Publishing},
	author = {Livan, Giacomo and Novaes, Marcel and Vivo, Pierpaolo},
	year = {2018},
	doi = {10.1007/978-3-319-70885-0},
}

@book{potters_first_2020,
	address = {Cambridge},
	title = {A {First} {Course} in {Random} {Matrix} {Theory}: for {Physicists}, {Engineers} and {Data} {Scientists}},
	isbn = {978-1-108-48808-2},
	shorttitle = {A {First} {Course} in {Random} {Matrix} {Theory}},
	url = {https://www.cambridge.org/core/books/first-course-in-random-matrix-theory/2292A554A9BB9E2A4697C35BCE920304},
	urldate = {2025-11-20},
	publisher = {Cambridge University Press},
	author = {Potters, Marc and Bouchaud, Jean-Philippe},
	year = {2020},
	doi = {10.1017/9781108768900},
}

@article{ahmadian_properties_2015,
	title = {Properties of networks with partially structured and partially random connectivity},
	volume = {91},
	url = {https://link.aps.org/doi/10.1103/PhysRevE.91.012820},
	doi = {10.1103/PhysRevE.91.012820},
	number = {1},
	urldate = {2025-11-20},
	journal = {Physical Review E},
	author = {Ahmadian, Yashar and Fumarola, Francesco and Miller, Kenneth D.},
	month = jan,
	year = {2015},
	note = {Publisher: American Physical Society},
	pages = {012820},
}

@article{kadmon2015PRX,
  title = {Transition to {{Chaos}} in {{Random Neuronal Networks}}},
  author = {Kadmon, Jonathan and Sompolinsky, Haim},
  year = {2015},
  month = nov,
  journal = {Physical Review X},
  volume = {5},
  number = {4},
  pages = {041030},
  publisher = {{American Physical Society}},
  doi = {10.1103/PhysRevX.5.041030},
  urldate = {2023-10-02}
}

@article{harris2023PRR,
  title = {Effect of Sparsity on Network Stability in Random Neural Networks Obeying {{Dale}}'s Law},
  author = {Harris, Isabelle D. and Meffin, Hamish and Burkitt, Anthony N. and Peterson, Andre D. H.},
  year = {2023},
  month = nov,
  journal = {Physical Review Research},
  volume = {5},
  number = {4},
  pages = {043132},
  publisher = {{American Physical Society}},
  doi = {10.1103/PhysRevResearch.5.043132},
  urldate = {2023-11-27}
}

@article{mastrogiuseppe2018N,
  title = {Linking {{Connectivity}}, {{Dynamics}}, and {{Computations}} in {{Low-Rank Recurrent Neural Networks}}},
  author = {Mastrogiuseppe, Francesca and Ostojic, Srdjan},
  year = {2018},
  month = aug,
  journal = {Neuron},
  volume = {99},
  number = {3},
  pages = {609-623.e29},
  issn = {0896-6273},
  doi = {10.1016/j.neuron.2018.07.003},
  urldate = {2023-10-02}
}

@article{mastrogiuseppe2017PCB,
  title = {Intrinsically-Generated Fluctuating Activity in Excitatory-Inhibitory Networks},
  author = {Mastrogiuseppe, Francesca and Ostojic, Srdjan},
  year = {2017},
  month = apr,
  journal = {PLOS Computational Biology},
  volume = {13},
  number = {4},
  pages = {e1005498},
  publisher = {{Public Library of Science}},
  issn = {1553-7358},
  doi = {10.1371/journal.pcbi.1005498},
  urldate = {2023-11-27},
  langid = {english}
}

@article{miller2012NC,
  title = {Mathematical {{Equivalence}} of {{Two Common Forms}} of {{Firing Rate Models}} of {{Neural Networks}}},
  author = {Miller, Kenneth D. and Fumarola, Francesco},
  date = {2012-01-01},
  journaltitle = {Neural Computation},
  shortjournal = {Neural Computation},
  volume = {24},
  number = {1},
  pages = {25--31},
  issn = {0899-7667},
  doi = {10.1162/NECO_a_00221},
  urldate = {2024-01-20}
}

@article{Rost2018BiologicalCybernetics,
author = {Rost, Thomas and Deger, Moritz and Nawrot, Martin},
year = {2018},
month = {04},
pages = {},
title = {Winnerless competition in clustered balanced networks: inhibitory assemblies do the trick},
volume = {112},
journal = {Biological Cybernetics},
doi = {10.1007/s00422-017-0737-7}
}

@article {Rostami2020biorxiv,
	author = {Vahid Rostami and Thomas Rost and Alexa Riehle and Sacha J. van Albada and Martin P. Nawrot},
	title = {Excitatory and inhibitory motor cortical clusters account for balance, variability, and task performance},
	elocation-id = {2020.02.27.968339},
	year = {2022},
	doi = {10.1101/2020.02.27.968339},
	publisher = {Cold Spring Harbor Laboratory},
	URL = {https://www.biorxiv.org/content/early/2022/02/08/2020.02.27.968339},
	journal = {bioRxiv}
}

@ARTICLE{Weidel2021frontiers,
AUTHOR={Weidel, Philipp and Duarte, Renato and Morrison, Abigail},   	 
TITLE={Unsupervised Learning and Clustered Connectivity Enhance Reinforcement Learning in Spiking Neural Networks},      	
JOURNAL={Frontiers in Computational Neuroscience},      	
VOLUME={15},           	
YEAR={2021},      	  
URL={https://www.frontiersin.org/articles/10.3389/fncom.2021.543872}, 
DOI={10.3389/fncom.2021.543872},      
ISSN={1662-5188},   
}

@article{LitwinKumar2012natneuro,
author = {Litwin-Kumar, Ashok and Doiron, Brent},
year = {2012},
month = {09},
pages = {1498-1505},
title = {Slow dynamics and high variability in balanced cortical networks with clustered connections},
volume = {15},
journal = {Nature Neuroscience},
doi = {10.1038/nn.3220}
}

@ARTICLE{Setareh2017frontiers,  
AUTHOR={Setareh, Hesam and Deger, Moritz and Petersen, Carl C. H. and Gerstner, Wulfram},   	 
TITLE={Cortical Dynamics in Presence of Assemblies of Densely Connected Weight-Hub Neurons},      	
JOURNAL={Frontiers in Computational Neuroscience},      	
VOLUME={11},           	
YEAR={2017},      
URL={https://www.frontiersin.org/articles/10.3389/fncom.2017.00052},	
DOI={10.3389/fncom.2017.00052},      
ISSN={1662-5188},   
}

@article{Rosenbaum2014PRX,
  title = {Balanced Networks of Spiking Neurons with Spatially Dependent Recurrent Connections},
  author = {Rosenbaum, Robert and Doiron, Brent},
  journal = {Phys. Rev. X},
  volume = {4},
  issue = {2},
  pages = {021039},
  numpages = {9},
  year = {2014},
  month = {May},
  publisher = {American Physical Society},
  doi = {10.1103/PhysRevX.4.021039},
  url = {https://link.aps.org/doi/10.1103/PhysRevX.4.021039}
}

@article{Rosenbaum2017NatNeuro,
author = {Rosenbaum, Robert and Smith, Matthew and Kohn, Adam and Rubin, Jonathan and Doiron, Brent},
year = {2016},
month = {10},
pages = {},
title = {The spatial structure of correlated neuronal variability},
volume = {20},
journal = {Nature neuroscience},
doi = {10.1038/nn.4433}
}

@article{znamenskiy2024functional,
  title={Functional specificity of recurrent inhibition in visual cortex},
  author={Znamenskiy, Petr and Kim, Mean-Hwan and Muir, Dylan R and Iacaruso, M Florencia and Hofer, Sonja B and Mrsic-Flogel, Thomas D},
  journal={Neuron},
  volume={112},
  number={6},
  pages={991--1000},
  year={2024},
  publisher={Elsevier}
}

@article{levy2012spatial,
  title={Spatial profile of excitatory and inhibitory synaptic connectivity in mouse primary auditory cortex},
  author={Levy, Robert B and Reyes, Alex D},
  journal={Journal of Neuroscience},
  volume={32},
  number={16},
  pages={5609--5619},
  year={2012},
  publisher={Soc Neuroscience}
}

@article{schneider2023cell,
  title={Cell-type-specific inhibitory circuitry from a connectomic census of mouse visual cortex},
  author={Schneider-Mizell, Casey M and Bodor, Agnes L and Brittain, Derrick and Buchanan, JoAnn and Bumbarger, Daniel J and Elabbady, Leila and Gamlin, Clare and Kapner, Daniel and Kinn, Sam and Mahalingam, Gayathri and others},
  journal={bioRxiv},
  year={2023},
  publisher={Cold Spring Harbor Laboratory Preprints}
}

@article{langton1990computation,
  title={Computation at the edge of chaos: Phase transitions and emergent computation},
  author={Langton, Chris G},
  journal={Physica D: nonlinear phenomena},
  volume={42},
  number={1-3},
  pages={12--37},
  year={1990},
  publisher={Elsevier}
}

@article{vanVreeswijk1996,
  title = {Chaos in {{Neuronal Networks}} with {{Balanced Excitatory}} and {{Inhibitory Activity}}},
  author = {van Vreeswijk, C. and  Sompolinsky, H.},
  date = {1996-12-06},
  journaltitle = {Science},
  volume = {274},
  number = {5293},
  eprint = {8939866},
  eprinttype = {pmid},
  pages = {1724--1726},
  publisher = {{American Association for the Advancement of Science}},
  issn = {0036-8075, 1095-9203},
  doi = {10.1126/science.274.5293.1724},
  urldate = {2020-10-26},
  langid = {english}
}

@book{binney_theory_2001,
	title = {The {Theory} of {Critical} {Phenomena}: an introduction to the renormalization group},
	isbn = {0-19-851393-3},
	publisher = {Oxford Univ. Press},
	author = {Binney, J. J. and Dowrick, N. J. and Fisher, A. J. and Newman, M. E. J.},
	year = {2001}
}

@article{Lorenz1963,
  author = {Lorenz, E. N.},
  title = {Deterministic nonperiodic flow},
  journal = {Journal of the Atmospheric Sciences},
  volume = {20},
  number = {2},
  pages = {130--141},
  year = {1963}
}

@article{Aizawa1982GlobalAO,
  title={Global Aspects of the Dissipative Dynamical Systems. I — Statistical Identification and Fractal Properties of the Lorenz Chaos —},
  author={Yoji Aizawa},
  journal={Progress of Theoretical Physics},
  year={1982},
  volume={68},
  pages={64-84},
  url={https://api.semanticscholar.org/CorpusID:124943762}
}

@article{Chen1999,
  author = {Chen, Guanrong and Ueta, Tetsushi},
  title = {Yet another chaotic attractor},
  journal = {International Journal of Bifurcation and Chaos},
  volume = {9},
  number = {7},
  pages = {1465--1466},
  year = {1999}
}

@inbook{Halvorsen,
title = {Halvorsen System},
booktitle = {Elegant Automation},
chapter = {Chapter 32},
pages = {197-202},
doi = {10.1142/9789811277528_0032},
URL = {https://www.worldscientific.com/doi/abs/10.1142/9789811277528_0032},
eprint = {https://www.worldscientific.com/doi/pdf/10.1142/9789811277528_0032}
}

@article{bertschinger2004real,
  title={Real-time computation at the edge of chaos in recurrent neural networks},
  author={Bertschinger, Nils and Natschl{\"a}ger, Thomas},
  journal={Neural computation},
  volume={16},
  number={7},
  pages={1413--1436},
  year={2004},
  publisher={MIT Press}
}

\end{document}


\title{Supplementary information}
\author{Emmanouil Giannakakis*, Victor Buend\'ia*,  Oleg Vinogradov, Anna Levina}

\maketitle
\tableofcontents

\section{Model} 	

We consider a system with $N$ neurons, organized in $M$ clusters. Each cluster has an excitatory and an inhibitory population, the total number of populations is $P=2M$. The $i$-th neuron of population $\mu$ is characterized by a continuous firing rate $x_i ^\mu>0$. Notice that $i=1,2,\ldots,N_\mu$, where $N_\mu$ is the number of neurons in population $\mu$, with $\mu = 1,2,\ldots,P$.  The equations for the dynamical state of the reservoirs read
\begin{equation}
    \dot x_i ^\mu = -x_i ^\mu + \phi \left( \sum_{\nu=1} ^{P} \sum_{j=1} ^{N_\nu} W_{ij} ^{\mu \nu} x_j ^\nu  \right),
\end{equation} 
where $\phi(x) = 1/(1+\exp(-x))$ is a sigmoidal transfer function and $\mat{W}$ the connectivity matrix with elements $W_{ij} ^{\mu\nu}$. 

Observe that we have selected our transfer function in such a way that $\phi(0) = 1/2$ and $\phi(x) > 0 \quad \forall x$, which ensures Dale's law by keeping all $x>0$ through the dynamics. 
The most common choice for nonlinearities in neural networks are those such that $\phi(x \leq 0) = 0$. However, this makes learning for the reservoir computer technically more cumbersome.  Our choice facilitates learning while respecting Dale's law.

The first step we will pursue is a change of variables, $\vec y = \mat{W} \vec x$. The dynamics of $y_j ^\mu (t)$ are 
\begin{equation}
    \dot y_i ^\mu = -y_i ^\mu + \sum_{\nu=1} ^{P} \sum_{j=1} ^{N_\nu} W_{ij} ^{\mu\nu} \phi(y_j ^\nu). 
    \label{eq:model_currents}
\end{equation}
The difference between the variables $x_i ^\mu$ and $y_i ^\mu$ is that of rate activity to a membrane potential, and their dynamics have been shown to be equivalent \cite{miller2012NC}. In our case, $y_i ^\mu$ represents the input current to the $i$-th neuron in population $\mu$.

\section{Baik-Ben Arous-Peché transition from clustered connectivity} 	

\subsubsection*{Clustered connectivity matrix and normalization}

\newcommand{\unw}{\mat{\tilde W}}

We will follow a standard approach to model the clustered connectivity. As the clusters lead to a matrix formed by blocks, we will approximate each block itself by a Gaussian matrix, a system which is analytically tractable.

Thus, we approximate
\begin{equation}
    \tilde W_{ij} ^{\mu\nu} \simeq w_{\mu\nu} + \sigma_{\mu\nu} z_{ij} ^{\mu\nu},
\end{equation}
which is valid as long as each block has well-defined mean $w_{\mu\nu}$ and variance $\sigma_{\mu\nu} ^2$. The next step is to find a proper normalization scheme. The dynamics of the neural network depend on the eigenvalues of $\unw$, which scales with the system size. In dynamical mean-field theory, the classical normalization procedure is
\begin{equation}
    W_{ij} ^{\mu\nu} \simeq \frac{w_{\mu\nu}}{N} + \frac{\sigma_{\mu\nu}}{\sqrt N} z_{ij} ^{\mu\nu},
\end{equation}
where the rationale for this normalization is the following: sums over $j$ yield $\mathcal O(N)$ terms when applied to the first term, but $\sum_j z_{ij} ^{\mu\nu} \sim O(\sqrt N)$. In this way, the matrix $\mat{W}$ is well-defined in the thermodynamic limit. 

\newcommand{\lot}{\lambda_\text{ot}}
\newcommand{\lsr}{\lambda_\text{sr}}

However, in reservoir computing it is desirable to keep control of the largest eigenvalue as a control parameter, $\lambda$. To do so, one customarily scales the matrix by $\lambda / \lambda_W$, being $\lambda_{\tilde W}$ the largest eigenvalue of $\mat{W}$. Therefore, we normalize as 
\begin{equation}
    W_{ij} ^{\mu\nu} \simeq \frac{\lambda}{\lambda_{\tilde W}} \tilde W_{ij} ^{\mu\nu},
    \label{eq:norm_w}
\end{equation}
The main question arising now is the finite-size scaling of $\lambda_{\tilde W}$. Our Gaussian expansion for $\unw$ can be thought as a low-rank perturbation to a Gaussian random matrix by blocks. The spectra of such matrices can be dominated either by low-rank outliers or by a continuum density of eigenvalues. The discrete spectral outlier of unnormalized $\unw$ is $\lot = N \lambda_w / P$, being $\lambda_w$ the largest eigenvalue of $\mat{w}$, while its spectral radius is given by $\lsr =  \lambda_\sigma \sqrt {N / P}$, with $\lambda_\sigma ^2$ the largest eigenvalue of $\mat{\sigma}$ \cite{potters_first_2020}.  

Thus, the largest eigenvalue $\lambda_{\tilde W} = \max (\lot, \lsr)$ scales with $N$ when the low-rank contribution dominates and with $\sqrt N$ when the contribution from the spectral disk dominates. We will refer to these cases as the low-rank-dominated phase or bulk-dominated phase, respectively. We will show that modulating the network's clustering can lead to a change of the dominant contribution, an instance of the so-called a Baik-Ben Arous-Peche critical transition in random matrix theory \cite{baik_phase_2005, potters_first_2020}. 

\subsubsection*{Structure of the reduced matrices}

\newcommand{\pin}[1]{p_{\text{in}, #1}}
\newcommand{\pout}[1]{p_{\text{out}, #1}}

We have seen that the spectra of the connectivity matrix $\unw$ depends only on the reduced matrices $\mat{w}$ and $\mat{\sigma}$. Therefore, the first step is to characterize them for our connectivity matrix.

The weight of each link in the reservoir is sampled from a uniform distribution $[w_0, w_0 + \Delta w]$. The mean and variance of such a uniform distribution are given by $w = w_0 + \Delta w / 2$ and $\sigma _w ^2 = (\Delta w)^2/12$, respectively. Neurons in the reservoir are connected with a probability $p_c$, so the resulting mean and variance for each block can be found by the law of total variance, giving $w_0 p_c$ for the mean and $p_c \left( \sigma _w ^2 + w_0^2 (1-p_c)  \right)$ for the variance. 

The probability of connection $p_c$ depends on whether the neurons belong to the same module. Neurons within the same module are connected with probability $\pin{X} = p(1 + (M-1)r_X)/M$, whereas neurons belonging to different modules are connected with probability $\pout{X} = p (1 - r_X) /M$. Here, $X$ denotes the kind of connection (e.g., excitatory-excitatory, inhibitory-excitatory).

Using $r_E$ and $r_I$, for the modularity of connections for excitatory and inhibitory connections, respectively, the reduced matrices read  
\begin{equation}
    w_{\mu \nu} = 
    \begin{cases}
        w_0 \pin{E},   &   \nu < M, \quad  \mu = \nu \mod M, \\
        w_0 \pout{E},  &   \nu < M, \quad  \mu \neq \nu \mod M, \\
        -w_0 \pin{I},  &  \nu > M,  \quad \mu = \nu \mod M, \\
        -w_0 \pout{I}, & \nu > M,   \quad  \mu \neq \nu \mod M,
    \end{cases} 
\end{equation}
for the means and 
\begin{equation}
    \sigma_{\mu \nu} ^2 =
    \begin{cases}
        \pin{E} \left( \sigma_w ^2 + w _0^2(1-\pin{E}) \right ),   & \nu < M, \quad  \mu = \nu \mod M, \\
        \pout{E} \left( \sigma_w ^2 + w_0^2(1-\pout{E}) \right ), & \nu < M, \quad  \mu \neq \nu \mod M, \\
        \pin{I} \left( \sigma_w ^2 + w_0^2(1-\pin{I}) \right ),   & \nu > M,  \quad \mu = \nu \mod M, \\
        \pout{I} \left( \sigma_w ^2 + w_0^2(1-\pout{I}) \right ), & \nu > M,   \quad  \mu \neq \nu \mod M,
    \end{cases} 
\end{equation}
for the variances.

\subsubsection*{BBP transition for for $M=2$}

For two modules, $M=2$, the eigenvalues of the reduced matrices $\mat{w}$ and $\mat{\sigma}$ can be easily computed, since they are the eigenvalues of $4\times4$ matrices. Thus, the eigenvalues of $\unw/N$ are 
\begin{subequations}
\begin{align}
    \lot =& \frac{1}{4}p (r_E-r_I)w_0, \\
    \lsr =&  \sqrt{ \frac{p}{8N}\left[  4p\sigma_w^2+pw_0^2 \left(4- p(2+r_E^2+r_I^2)   \right)   \right]},
\end{align}
\end{subequations}
where we have already set $P=4$ explicitly. Observe that although $\mat{W}$ is balanced in the thermodynamic limit (since $\sum _\nu w_{\mu \nu} = 0$) the outlier can get large and positive due to the structure of the network. In contrast, a perfectly balanced network will always have $\lambda_\text{ot} = 0$. The isolated outlier will be responsible for a saddle-node bifurcation in the dynamics. For chaos to happen instead, the spectral disk must cross the instability threshold. Therefore, a necessary condition for chaos is that the outlier $\lambda_\text{ot}$ must be to the right, outside the spectral radius. This happens when the outlier enters the spectral disk, at $\lot = \lsr$. The equality leads to a critical boundary for fixed inhibitory clustering, 
\begin{equation}
    r_E ^c = \frac{N}{(N+2)} r_I +  \frac{2}{(N+2)pw_0^2}\sqrt{(N+2)\left(2\sigma_{w}^{2}+\left(2-p\right)w_{0}^{2}\right)-pw_{0}^{2}(N+1)r_{I}^{2}}.
\end{equation}
This expression is very close to a straight line in the interval $r_I\in[0,1]$. Taking also $N\gg1$, we get 
\begin{equation}
    r_E ^c \simeq r_I + \frac{2\sqrt{ 2\sigma_w^2+(2-p)w_0^2 } }{w_0\sqrt{pN}}.
    \label{eq:rec}
\end{equation}
We keep the order $\mathcal{O}(N^{-1/2})$ as our reservoirs have a finite size. The bifurcation described by $r_e^c$ is plotted in the Main Text, Fig. 3c. Since this simply separates the space in two regions with identical dynamics, we restrict the clustering parameters to the line $r_E=r,r_I=1-r$ without loss of generality. When this is done, the eigenvalues simplify to 
\begin{subequations}
\begin{align}
    \lot =& \frac{1}{4}p (2r-1)w_0, \\
    \lsr =&  \sqrt{\frac{p}{8N} \left[ 4\sigma_w^2 + w_0 ^2 \left(4 - p (3-2r(1-r)) \right) \right]  },
\end{align}
\end{subequations}
and the exact expression for the critical value at which both eigenvalues coincide is given by 
\begin{equation}
    r_c = \frac{1}{2} +  \frac{1}{2w_0} \sqrt{\frac{8(\sigma_w^2 + w_0^2) -5pw_0^2}{p(1+N)}}. 
    \label{eqs:rc}
\end{equation}
We will use simply $r$ as a control parameter from now on. It is important to remark that although $\lot$ becomes negative at $r=1/2$, when both excitation and inhibition populations are equally clustered, the outlier blends with the spectral disk before. At a finite size, this leads to an "asymmetry" in the location of the bifurcation, benefitting a structure where E is clustered and I homogeneous.

Observe that the value of the critical point does not depend on our normalization choice for $\unw$, as we precisely impose $\lot = \lsr$. Interestingly, if the usual DMFT normalization scheme is considered instead, this extra $\sqrt N$ is cancelled and then $r_c$ does not scale with the system size, leading to an asymmetry between excitation and inhibition modularity even in the thermodynamic limit. 

\begin{figure*}
    \includegraphics[width=\linewidth]{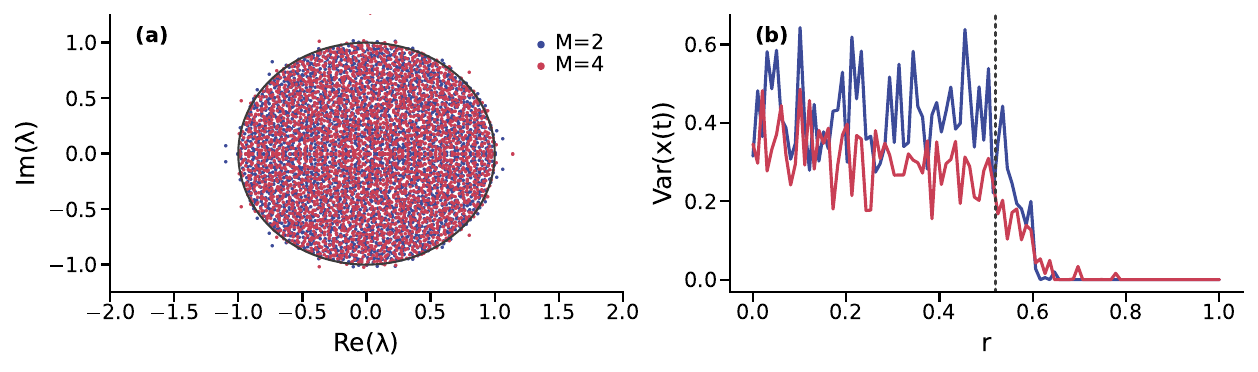}
    \caption{\textbf{Network's dynamical properties are independent of number of modules (a)} Spectrum two networks with 2 and 4 modules and $N=3000$, normalized to have unit spectral radius. (b) Variance of the individual rate, which indicates a chaotic transition for fixed spectral radius $\lambda = 10$ and increasing modularity for the two previous networks. The transition to chaos happens at $r_c$ (eq. \ref{eqs:rc}), which is the value computed for $M=2$.}
    \label{fig:equivalence_clusters}
\end{figure*}

If we scale $\unw$ by a control parameter $\lambda$, $r_c$ remains invariant. As we will show in the next section, when the largest eigenvalue $\lambda$ is large enough, $r_c$ becomes the \emph{edge of chaos}.  Interestingly, the value $r_c$ obtained for $M=2$ seems to be a good approximation for larger values of $M$, such as $M=4$ used in the main text. A phase diagram as a function of $r_c$ for different numbers of modules can be seen in Sup. Fig. \ref{fig:equivalence_clusters}.

Another feature of Sup. Fig. \ref{fig:equivalence_clusters} that is worth mentioning is the presence of outliers outside the spectral disk in the bulk-dominated phase. Commonly these arise from non-Gaussian contributions to the network or finite-size effects. However, here they are produced by our normalization scheme plus the lack of exact balance in the microscopic model. The input current is proportional to the sum of columns of the connectivity matrix, which in turn is proportional to $\sum_\nu w_{\mu \nu}$. In theory, our network is balanced, $\sum_\nu w_{\mu \nu} = 0$, preventing any low-rank mode to emerge. In practice, samples from the microscopic model display fluctuations. In the bulk-dominated regime, the low-rank term actually scales as $\sqrt N w_{\mu \nu}$, meaning that fluctuations are $\mathcal O(1)$. A sampled network will have $\sum_\nu w_{\mu\nu} \neq0$, breaking exact balance and leading to emergent outliers even at the thermodynamic limit. These would never appear in the usual DMFT normalization, in which the low-rank contribution does not scale with $N$.

\section{Dynamic Mean-Field Theory} 

This section contains the detailed analytical calculations to obtain the phase diagram of the system with respect to the maximum eigenvalue $\lambda$ and the critical point locations. 

We start substituting the expression for the normalized connectivity Eq. (\ref{eq:norm_w}) in the model equation for input currents, Eq. (\ref{eq:model_currents}), 
\begin{equation}
    \dot y_i ^\mu = -y_i ^\mu + \frac{\lambda}{\lambda_{\tilde W}} \sum_{\nu=1} ^{P} w_{\mu\nu} \sum_{j=1} ^{N_\nu} \phi(y_j ^\nu) + 
    \frac{\lambda}{\lambda_{\tilde W}} \sum_{\nu=1} ^{P} \sigma_{\mu\nu} \sum_{j=1} ^{N_\nu} z_{ij} ^{\mu\nu} \phi(y_j ^\nu).
\end{equation}
The main idea is to assume that the sums over $j$ can be approximated by random Gaussian variables. When this is done, index $i$ becomes dummy, as interactions between neurons become a single unit subject to a stochastic (mean) field. Removing index $i$, we define two random variables,
\begin{subequations}
\begin{align}
    m_\mu (t)     &= \frac{1}{N} \sum_j ^{N_\mu} \phi(y_j ^\mu (t)), \\
    s_{\mu\nu} (t) &= \frac{1}{\sqrt N} \sum_j ^{N_\mu} z_j ^{\mu\nu}\phi(y_j ^\nu (t)).
\end{align}
\end{subequations}
Then, we can average over the disorder to obtain the average of these new variables. We employ a typical assumption in DMFT: correlation between $z_j ^{\mu\nu}$ and any function of $y^\mu$, as well as correlations among different units, is of order $1/N$ and can be neglected in the thermodynamic limit.
Then, 
\begin{subequations}
\begin{align}
    \langle m_\mu \rangle      &= a_\mu \langle \phi(y_\mu) \rangle, \\
    \langle s_{\mu\nu} \rangle &= 0,
\end{align}
\end{subequations}
where we have used $\langle \phi(y_j ^\mu) \rangle =\langle \phi(y ^\mu) \rangle \equiv \langle \phi(y _\mu) \rangle  $, as the average over the disorder eliminates the dependence on $j$, and $\langle z_j ^{\mu\nu}  \rangle = 0$.  Additionally, we have defined the ratio of neurons in population $\mu$, $a_\mu = N_\mu / N = 1 / P$ in our case. 

Then we compute the variance of both variables. Observe that all other covariances go to zero, due to assumption that correlations between units vanish in the thermodynamic limit. The assumption allows us to simplify the calculation by breaking the double sum into two parts, as 
\begin{align}
    \langle m_\mu ^2 \rangle - \langle m_\mu \rangle ^2  &=
    \frac{1}{N^2}  \sum_{j,k} \left[ \langle \phi(y_j ^\mu) \phi(y_k ^\mu) \rangle - \langle \phi(y_j ^\mu) \rangle \langle \phi(y_k ^\mu) \rangle \right] = \nonumber \\
    &= \frac{1}{N^2} \left( \sum_{j\neq k } (\ldots) + \sum_j ^{N_\mu} \left[ \langle \phi(y_j ^\mu) ^2 \rangle - \langle \phi(y_j ^\mu) \rangle ^2 \right]  \right) = \nonumber \\
    &= \frac{a_\mu }{N}  \left[ \langle \phi(y _\mu) ^2 \rangle - \langle \phi(y_\mu) \rangle^2  \right] \to 0.
\end{align}
Again, we have taken out the averages from the sum to discover that fluctuations of $m_\mu$ vanish in the thermodynamic limit. Therefore, $m_\mu$ converges to its mean in the thermodynamic limit, and can be considered a scalar with value $m_\mu (t) = a_\mu \langle \phi(y_\mu (t)) \rangle$

Note that we evaluated both $m_\mu(t)$ at the same time $t$, but it is immediate to see that the same result stands for two different time points $t, t'$. 

For $s_{\mu\nu}$, however, it is important to make the time dependencies explicit, as

\begin{align}
    \langle s_{\mu\nu}(t) s_{\mu\nu}(t') \rangle &= \frac{1}{N} \sum_{j,k} \langle z_j ^{\mu\nu} z_k ^{\mu\nu} \rangle \langle \phi(y_j ^\nu(t)) \phi(y_k ^\nu(t')) \rangle = \nonumber \\
    &= \frac{1}{N} \sum_j ^{N_\nu} \langle \phi(y_j ^\nu(t)) \phi(y_j ^\nu(t')) \rangle
    = a_\nu \Delta ^{\nu\nu}_\phi (t, t'), 
    \label{eq:autocorr_s}
\end{align}
where we defined $\Delta^{\nu\nu}_\phi (t,t') =\langle \phi(y ^\nu(t)) \phi(y ^\nu(t')) \rangle$, the autocorrelation function of the nonlinearity. Finally,  the mean-field equations read 
\begin{equation}
    \dot y _\mu (t) = -y_\mu (t) + \frac{\lambda N}{\lambda_{\tilde W}} \sum_{\nu=1} ^{P} w_{\mu\nu} m_\nu (t) + 
    \frac{\lambda \sqrt N}{\lambda_{\tilde W}} \sum_{\nu=1} ^{P} \sigma_{\mu\nu} s_{\mu\nu} (t),
    \label{eq:fulldmft}
\end{equation}
and $s_{\mu\nu}(t)$ is a Gaussian colored noise with zero-mean and autocorrelation function given by Eq. (\ref{eq:autocorr_s}).

\subsection{Reservoir normalization}

\subsubsection*{Low-rank-dominated regime}

To study the behaviour of Eq. (\ref{eq:fulldmft}), we just need to set the scaling of $\lambda_{\tilde W}$. Let's start with the low-rank-dominated regime. In this case, we had $\lambda_{\tilde W} = N \lambda_w / P$. Therefore, the second term of Eq. (\ref{eq:fulldmft}), vanishes in the thermodynamic limit. Since  $m_\nu(t)$ does not fluctuate, in this case we have a fully deterministic evolution, 
\begin{equation}
    \dot y _\mu (t) = -y_\mu (t) + \frac{\lambda}{\lambda_w} \sum_{\nu=1} ^{P} w_{\mu\nu} \phi(y_\mu (t)).
\end{equation}

Observe that we substituted $a_\nu = 1/P$, which was simplified, and we wrote $\langle \phi(y_\mu) \rangle = \phi(y_\mu)$, as there is no noise in this case. It is a $P$-dimensional dynamical system.

To find fixed points, we write our nonlinearity as $\phi(x) = \bar \phi(x) + \phi_0$, with $\phi_0 \equiv \phi(0)$. Then, since our network is balanced, it turns out that $\sum_\nu w_{\mu\nu} \phi(y_\mu) = \sum_\nu w_{\mu\nu} \bar \phi(y_\mu)$. Then one can check that $y_\mu = 0$ is a fixed point of the dynamical system. 

The Jacobian matrix of this system can be computed as usual, yielding 
\begin{equation}
    J_{\mu\nu}(\{ y_\mu \}) = -\delta_{\mu\nu} + \frac{\lambda}{\lambda_w} w_{\mu \nu} \phi'(y_\nu). 
\end{equation}

Thus, the Jacobian at the fixed point is $\mat{J} = -\mat{I} + \lambda \phi'(0) \mat{w} / \lambda_w$. Its eigenvectors are those of $\mat{w}$, and the eigenvalues are $-1 + \lambda \phi'(0)$, leading to a critical point
\begin{equation}
    \lambda_c ^{lr} = \frac{1}{\phi'(0)}
\end{equation}

\subsubsection*{Bulk-dominated regime}

This case is more complicated. The system has heterogeneity again, as $s_{\mu\nu}$ is a fluctuating variable, and thus $y_\mu$ is given by a (multidimensional) distribution of fixed points rather than a scalar. Moreover, the first term now scales badly with $\sqrt N$,
\begin{equation}
    \dot y _\mu (t) = -y_\mu (t) + \frac{\sqrt N\lambda}{\lambda_w} \sum_{\nu=1} ^{P} w_{\mu\nu} \langle\bar  \phi(y_\mu (t)) \rangle + \frac{\sqrt P \lambda}{\lambda_\sigma} \sum_{\nu=1} ^{P} \sigma_{\mu\nu} s_{\mu\nu}(t).
    \label{eq:bulkeq}
\end{equation}
The $\sqrt N$ term must vanish for the system to be well-defined. Since $\bar \phi(x)$ is an odd sigmoid function, a symmetric distribution of fixed points around the origin will cancel this term. Hence, the solution must fulfill $\langle y_\mu(t) \rangle = 0$.

Once we have cancelled the low-rank term, fixed points satisfy
\begin{equation}
y_\mu  = \frac{\sqrt P \lambda}{\lambda_\sigma} \sum_{\nu=1} ^{P} \sigma_{\mu\nu} s_{\mu\nu}.
\end{equation}
Since $s_{\mu\nu}$ is a Gaussian random variable, $y_\mu$ is also Gaussian. Its variance is
\begin{align}
    \langle y_\mu ^2 \rangle  &=  \frac{P \lambda^2}{\lambda_\sigma^2} \sum_{\alpha, \beta} ^{P} \sigma_{\mu\alpha} \sigma_{\mu\beta} \langle s_{\mu\alpha} s_{\mu\beta} \rangle =
    \frac{P \lambda^2}{\lambda_\sigma^2} \sum_{\alpha} ^{P} \sigma_{\mu\alpha} ^2 a_\alpha \Delta^{\alpha \alpha}_\phi
    = \frac{\lambda^2}{\lambda_\sigma^2} \sum_{\alpha} ^{P} \sigma_{\mu\alpha} ^2  \langle \phi(y_\alpha)^2 \rangle.
    \label{eq:varianceS}
\end{align}
Therefore, the distribution of fixed points is a centered multivariate Gaussian, where the covariance matrix is diagonal. Finally, the stability of this distribution remains to be studied.

To do so, the Jacobian must be computed. The complicated part is applying the derivative to the random variable $s_{\mu\nu}$. The simplest approach is to compute a perturbation to $s_{\mu\nu}(t)$ from the fixed point:
\begin{align}
    s_{\mu\nu}(t) &= s_{\mu\nu} + \delta s_{\mu\nu}(t) = \frac{1}{\sqrt N} \sum_j ^{N_\nu} z_j ^{\mu\nu}  \phi(y_j ^\nu + \varepsilon_j ^\nu (t)) \nonumber \\  
    &= \frac{1}{\sqrt N} \sum_j ^{N_\nu} z_j ^{\mu\nu} \left[ \phi(y_j ^\nu) +  \phi'(y_j^\nu) \varepsilon_j ^\nu (t) \right]  \nonumber \\
    &= s_{\mu\nu} + \frac{1}{\sqrt N} \sum_j ^{N_\nu} z_j ^{\mu\nu} \phi'(y_j^\nu) \varepsilon_j ^\nu (t).
\end{align}
This leads immediately to the definition of 
\begin{equation}
    \delta s_{\mu\nu}(t) = \frac{1}{\sqrt N} \sum_j ^{N_\nu} z_j ^{\mu\nu} \phi'(y_j^\nu) \varepsilon_j ^\nu
    \label{eq:delta_s}
\end{equation}
as a new random Gaussian variable. Its mean and variance can be computed exactly as we did for $s_{\mu\nu}(t)$, leading to 
\begin{subequations}
\begin{align}
    \langle \delta s_{\mu\nu}(t) \rangle &= 0,\\
    \langle \delta s_{\mu\nu}(t) \delta s_{\mu\nu}(t') \rangle &= a_\nu \langle \phi'(y_\nu) ^2 \rangle \Delta^{\nu\nu}_\varepsilon(t,t') \equiv 
    a_\nu \langle \phi'(y_\nu) ^2 \rangle C_\nu(t,t'),
\end{align}
\end{subequations}
where we have defined $C_\nu(t,t')$ to lighten the notation. Thus, $\delta s_{\mu\nu}(t)$ can be also interpreted as a Gaussian colored noise. If we now expand $y_\mu(t) = y_\mu + \varepsilon_\mu(t)$ in Eq. (\ref{eq:bulkeq}) the evolution for the perturbations $\varepsilon_\mu(t)$ is
\begin{equation}
    \dot \varepsilon _\mu (t) = -\varepsilon_\mu (t) + \frac{\sqrt P \lambda}{\lambda_\sigma} \sum_{\nu=1} ^{P} \sigma_{\mu\nu} \delta s_{\mu\nu}(t).
    \label{eq:perturbations}
\end{equation}

Taking expectation values, we have $\langle \dot \varepsilon _\mu \rangle = -\langle \varepsilon_\mu \rangle$, so in average all perturbations will decay to the fixed point at infinite time. However, the autocorrelation function of the noise depends itself on the autocorrelation of the variables, $C_\mu(t,t')$. 

We can obtain an equation for the autocorrelation function by isolating the noise in the right-hand side, multiplying both sides at different time points, and then taking averages. Thus, we have  
\begin{align}
    (\partial_t + 1)(\partial_{t'} + 1)C_\mu(t,t') &= \frac{P \lambda^2}{\lambda_\sigma^2} \sum_{\alpha, \beta} ^{P} \sigma_{\mu\alpha} \sigma_{\mu\beta} \langle \delta s_{\mu\alpha}(t) \delta s_{\mu\beta}(t') \rangle \nonumber \\
    &= \frac{\lambda^2}{\lambda_\sigma^2} \sum_{\alpha, \beta} ^{P} \sigma_{\mu\alpha} ^2  \langle \phi'(y_\alpha) ^2 \rangle C_\alpha(t,t'),
\end{align}
which is obtained analogously to Eq. (\ref{eq:varianceS}). 

We are looking for stationary state solutions. Therefore, letting $\tau = t-t'$ and assuming $C(t,t') = C(\tau)$, we have $\partial_t = \partial_\tau$, $\partial_{t'} = -\partial_\tau$ so our expression simplifies to 
\begin{align}
    \ddot C_\mu(\tau) &= C_\mu(\tau) - \frac{\lambda^2}{\lambda_\sigma^2} \sum_{\nu=1} ^{P} \sigma_{\mu\nu} ^2 \langle \phi'(y_\nu) ^2 \rangle C_\nu(\tau),
\end{align}
which can be cast into matrix form, 
\begin{equation}
    \ddot {\vec C} (\tau) = \left( -\mat{I} + \frac{\lambda^2}{\lambda_\sigma ^2} \mat{S} \right) \vec C(\tau). 
\end{equation}
The elements of matrix $\mat{S}$ are given by $S_{\mu\nu} = \sigma_{\mu\nu} ^2 \langle \phi'(y_\nu) ^2 \rangle$. This system has a phase transition when the largest eigenvalue of the linear coefficient changes sign. Its eigenvalues are those of $\mat{S}$. Therefore, the transition happens at 
\begin{equation}
    \lambda_c ^b = \sqrt \frac{\lambda_\sigma ^2}{\lambda_S}
    \label{eq:general_bulk_crit}.
\end{equation}

\subsubsection*{Approximation of critical point in the bulk-dominated regime}

Our DMFT analysis reveals that in the bulk-dominated regime, the fixed points follow a distribution with zero mean and finite variance $\langle y_\mu ^2 \rangle$ (Eq. (\ref{eq:varianceS})). This point loses stability at $\lambda_c ^b$, given by Eq. (\ref{eq:general_bulk_crit}), which depends on an expected value over the fixed points distribution.

For an exact solution, the variance of the fixed points needs to be found numerically, as Eq. (\ref{eq:varianceS}) is a self-consistent equation: the right-hand side contains an expectation value over the distribution of fixed points, which depends on the variance we are trying to find. Finally, the integral present in Eq. (\ref{eq:general_bulk_crit}) cannot be analytically evaluated. 

We present now an approximation to compute the variance and the location of the critical point. To start with, we assume that all modules are equivalent, and hence $
\langle y_\mu ^2\rangle, \langle\phi(y_\mu) \rangle, \langle \phi'(y_\mu)^2 \rangle$ do not depend on $\mu$. This ansatz is justified because our network is balanced by construction, so the input to each population $\sum _\nu w_{\mu\nu} \langle \phi(y_\nu)\rangle = \langle \phi(y)\rangle\sum _\nu w_{\mu\nu}=0$ becomes automatically balanced. 

Under this assumption, the determination of the variance reduces to 
\begin{equation}
    \langle y ^2 \rangle = \lambda^2 \langle \phi(y) ^2 \rangle,
    \label{eq:vari_aprox}
\end{equation}
as one can show that $\sum_\alpha \sigma_{\mu\alpha}^2 = \lambda_\sigma^2$. If the variance is very low, then the input is always close to zero and $\phi(y)$ can be expanded in Taylor series. The lowest order would be $\phi(0)^2 \simeq 1/4$. On the other hand, if the variance was very large, most of the sigmoid is saturating, meaning that it can be approximated by a step function. The average of the square of step function is $\langle \phi(y) \rangle \simeq 1/2$. Therefore, 
\begin{equation}
    \frac{\lambda^2}{4} \leq \Delta^2 \leq \frac{\lambda^2}{2}.
\end{equation}

For Eq. (\ref{eq:general_bulk_crit}), our assumption leads to  $\mat{S} = \langle \phi'(y)^2 \rangle \mat{\sigma}^2$ and therefore $\lambda_S = \langle \phi'(y)^2 \lambda_\sigma ^2$, so the criticality condition reduces to 
\begin{equation}
    \lambda^2 \langle \phi'(y)^2 \rangle = 1.
    \label{eq:crit_aprox}
\end{equation}

Now, the nonlinearity $\phi'(y)^2$ cannot be treated so easily. There are two options: first, assuming a small input current and expanding it in Taylor series. In this case,
we would find the critical threshold to be at least around $\lambda_c \simeq 4$, leading to large $\Delta$ and finding an inconsistent result. Therefore, a Taylor expansion will perform poorly. Another option is an asymptotic expansion for large variance,
\begin{equation}
    \frac{1}{\sqrt{2\pi\Delta^2}}e^{-y^2/(2\Delta^2)} \simeq \frac{1}{\sqrt{2\pi\Delta^2}} \left[ 1 - \frac{y^2}{2\Delta^2} + \mathcal(\Delta^{-4}) \right].
\end{equation}
At zeroth order the integral becomes simply 
\begin{equation}
    \langle \phi'(y)^2 \rangle \simeq \int \frac{dy}{\sqrt{2\pi\Delta^2}} \phi'(y)^2 = \frac{1}{6\sqrt{2\pi\Delta^2}}.
\end{equation}
Substituting the value of variance $\Delta^2 = \lambda^2/2$ and we can solve for the critical  $\lambda_c \simeq 6\sqrt\pi = 10.63$. Numerical solution of Eqs. (\ref{eq:vari_aprox}, \ref{eq:crit_aprox}) leads to an exact value $\lambda_c = 10.18$, so our approximation is quite accurate.

\subsection{Usual DMFT normalization}

In dynamical mean-field theory, the usual decomposition for the normalized connectivity matrix $\mat{W}$ is 
\begin{equation}
    W_{ij} ^{\mu\nu} = \frac{\lambda w_{\mu\nu}}{N} + \frac{\lambda \sigma_{\mu\nu}}{\sqrt N} z_{ij}^{\mu\nu}.
    \label{eq:dmft_norm}
\end{equation}
In this case, the outlier eigenvalue is given by $\lambda_\text{ot} = \lambda_w /P$, while the spectral radius is $\lambda_\text{sr} = \lambda_\sigma / \sqrt P$. All dependency with $N$ has been cancelled, as each term is scaled by its corresponding $N$. Therefore, the condition $\lambda_\text{ot}=\lambda_\text{sr}$ leads to a critical point $r_c \neq r_c(N)$, in contrast with the previous case,
\begin{equation}
    r_c = \frac{1}{2} + \frac{1}{2w_0} \sqrt{\frac{8\sigma_w^2+(8-5p)w_0^2}{1+p}}.
\end{equation}
We still have $r_c > 1/2$, so the asymmetry between excitatory and inhibitory modularity persists in the thermodynamic limit. There will still be a low-rank-dominated regime and a bulk-regime, so qualitatively the results are identical to the previous ones. 

We will just sketch the DMFT analysis, since this case is very similar to the previous computations, and it can be found also in the literature \cite{kadmon2015PRX, mastrogiuseppe2017PCB}. Inserting the connectivity into Eq.~(\ref{eq:model_currents}) and identifying the variables $m_\mu(t)$ and $s_{\mu\nu}(t)$ we arrive to 
\begin{equation}
    \dot y_\mu = -y_\mu + \lambda\sum_{\nu=1} ^{P} w_{\mu\nu} m_\nu(t) + \lambda\sum_{\nu=1} ^{P} \sigma_{\mu\nu} s_{\mu\nu}(t).
\end{equation}
Now, in contrast with the previous case, both terms are of the same order and contribute equally to the dynamics. Again, $m_\mu(t)$ behaves like a scalar, while $s_{\mu\nu}(t)$ is a colored white noise. Then, $y_\mu(t)$ must follow Gaussian statistics, as it is a combination of Gaussian variables. By taking averages, using that $\langle s_{\mu\nu} (t) \rangle = 0$ and $\langle m_\nu\rangle = \langle\phi (y_\nu) \rangle$, we get 
\begin{equation}
    \langle \dot y_\mu \rangle = -\langle y_\mu \rangle + \lambda\sum_{\nu=1} ^{P} w_{\mu\nu} \langle \bar \phi(y_\nu) \rangle 
\end{equation}
where we have used again that the network is balanced to interchange $\phi(y_\nu)$ by $\bar \phi(y_\nu)$. Let's start with the distribution of fixed points. If $\langle \dot y_\mu\rangle=0$, a distribution of fixed points with vanishing mean fulfills the equation. This is because $y_\mu$ must follow Gaussian statistics, as previously discussed, so if $\langle y_\mu\rangle=0$ then the distribution of fixed points is symmetric around the origin and $\langle \bar \phi(y_\nu) \rangle = 0$ automatically. Although there might be other solutions, we will consider first this one.   

The variance of the fixed points is given by 
\begin{equation}
    \langle y_\mu ^2 \rangle = \sum_{\alpha\beta} w_{\mu\alpha}w_{\mu\beta} \langle m_\alpha m_\beta \rangle +  \lambda^2 \sum_{\alpha}a_{\alpha}\sigma_{\mu\alpha}^{2}\langle \phi^2 (y_\alpha)\rangle,
\end{equation}
but $\langle m_\alpha m_\beta \rangle = \delta_{\alpha\beta}\langle m_\alpha ^2\rangle \to0$ in the thermodynamic limit, so the first term vanishes. One can observe that in this case, the variance of the monostable fixed will not change dramatically with $r$ as in the previous case. This can be understood intuitively because now both the outlier and bulk scale at the same order with the system, so the bulk can always contribute to the variance. 

In fact, in this case the low-rank dominated regime also experiences fluctuations, so even if a single outlier becomes unstable the system will not be fully deterministic. Here, the perturbations $y_\mu (t) =y_\mu +\varepsilon_\mu(t)$ lead to 
\begin{equation}
    \dot \varepsilon_\mu = -\varepsilon_\mu + \lambda\sum_{\nu=1} ^{P} w_{\mu\nu} \delta m_\nu(t) + \lambda\sum_{\nu=1} ^{P} \sigma_{\mu\nu} \delta s_{\mu\nu}(t),
\end{equation}
where $\delta s_{\mu\nu}$ is defined by Eq.~(\ref{eq:delta_s}) and $\delta m_\mu(t)$ is defined analogously by 
\begin{equation}
  \delta m_\mu (t)= \frac{1}{N} \sum_{j=1} ^{N_\mu} \phi'(y_j ^\mu) \varepsilon_j ^\mu(t).   
\end{equation}

As $m_\mu(t)$, $\delta m_\mu(t)$ converges to its average in the thermodynamic limit and behaves as a scalar. The perturbations $\varepsilon_\mu$ can become unstable in two ways. First, it might happen that its average its unstable, as 
\begin{equation}
    \langle \dot \varepsilon_\mu \rangle = -\langle \varepsilon_\mu \rangle + \frac{\lambda}{P}\sum_{\nu=1} ^{P} w_{\mu\nu} \langle\phi'(y_\nu) \rangle \langle\varepsilon_\nu\rangle,
\end{equation}
i.e., when the largest eigenvalue $\lambda_c^{lr} > 1 /\lambda_J$, being $\lambda_J$ the largest eigenvalue of the matrix given by elements $J_{\mu\nu} = w_{\mu\nu} \langle\phi'(y_\nu) \rangle/P $. Observe that this is almost the same result we got previously in the low-rank dominated regime. With the reservoir normalization, the variance of the fixed point vanishes, and so $\langle\phi'(y_\nu) \rangle \to \phi'(0)$. Additionally, we divided by the eigenvalue of $\lambda_w$. Therefore, this is the generalization of that formula for the case in which the system is subject to fluctuations.  

The other way the monostable fixed point can be become unstable is because, despite the average $\langle \varepsilon_\mu(t)\rangle$ might shrink to 0, the autocorrelations might be large. This leads to the same computation we performed already in the bulk-dominated case for the reservoir, with practically no difference. We will find a value $\lambda_c^b$ for which the autocorrelations change their behaviour, which depends again on the matrix $S_{\mu\nu} = \sigma_{\mu\nu}^2 \langle \phi'(y_\nu)^2\rangle/P$.

Observe that the first transition requires a single mode to become unstable, while the second requires a whole bulk of eigenvalues to occur. Therefore, which transition takes place depends on modularity $r$. Then, we have the same result as in the reservoir normalization. The major difference is that $\lambda_c^{lr}$ may change dramatically its value with respect to the previous one. However, qualitatively the phase diagram is identical.

\section{Dynamical state in $r_E$, $r_I$ phase space}

We have looked at the parameter space with $r_E = r$, $r_I = 1 - r$. Here, we show that the full $(r_E, r_I)$ does not display any qualitative difference, i.e. different dynamical regimes than the ones discussed in the Main Text. To show this, we run the dynamics of the recurrent network 100 times for each value of $r_E$ and $r_I$ and check that the results are symmetric to the diagonal of interest.

\begin{figure*}
    \includegraphics[width=\linewidth]{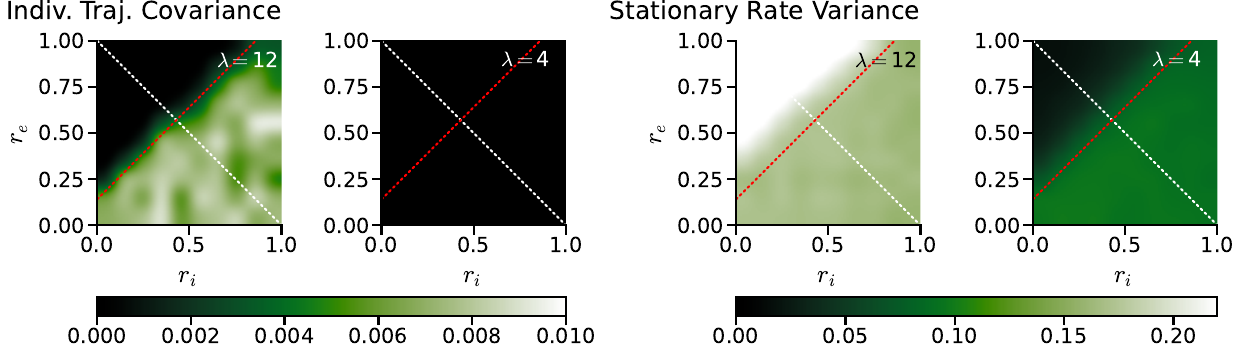}
    \caption{\textbf{Phase space for $(r_E, r_I)$.} (a) Average variance of the individual trajectories of neurons in the network for two spectral radii. The large one (left) shows a transition to chaos, displayed by the increase in each trajectory's variance. The smaller one (right) has no transitions. (b) Variance of the stationary rate values for the network. Both cases display a transition. White dashed line shows $r_E = 1-r_I$ line used in the main text. Red dashed line shows Eq.~(\ref{eq:rec}), which correctly predicts the transition to chaos in panel (a).}
    \label{fig:diagonal_dynamical}
\end{figure*}

In order to detect different states, we obtain the average variance of individual trajectories, as well as the variance of the stationary firing rates across all neurons for two different spectral radii. The results are shown in Fig. ~\ref{fig:diagonal_dynamical}. Observe that the bifurcation present in $(r_E, r_I)$ space is well predicted by Eq. ~(\ref{eq:rec}) in the chaotic case, as expected. 

For small values of the spectral radius, there's no chaotic state (average variance of trajectories is always low), and a transition from monostable to bistable can be found, hinted by the change in rate's variance. On the other hand, with high spectral radius the transition to chaos is visible by an increase of single trajectory variance. Variance across system's rates is not so informative, as this measure assumes rates are all stationary --which is false for the chaotic case-- but the transition is seen due to increase of spread of rate's mean. 

\section{Attractors}

We use the following 3-dimensional chaotic attractors as input time series (individually or all of them simultaneously) to our network \cite{Lorenz1963,Chen1999,Halvorsen,Aizawa1982GlobalAO}:

\hspace{1cm}

\textbf{Lorenz:} 
\begin{subequations}
\begin{align}
        \dot{x} &= \sigma (y - x), \\
        \dot{y} &= \rho x - y - xz,\\
        \dot{z} &= xy - \beta z,  
\end{align}
\end{subequations}
for $\sigma = 10$,  $\rho = 28$ and $\beta = 2.667$ 

\hspace{1cm}

\textbf{Aizawa:}
\begin{subequations}
\begin{align}
        \dot{x} &= (z - b)x - y, \\
        \dot{y} &= x + (z-b)y,\\
        \dot{z} &= c + az - z^3/3 - (x^2 + y^2)(1+ez) + fzx^3,
\end{align}
\end{subequations}
for $a = 0.95$, $b = 0.7$, $c = 0.6$, , $d = 3.5$, $e = 0.25$ and $f = 0.1$

\hspace{1cm}

\textbf{Halvorsen:}
\begin{subequations}
\begin{align}
        \dot{x} &= -\alpha x - 4y - 4z - y^2, \\
        \dot{y} &= -\alpha y - 4x - 4z - z^2,\\
        \dot{z} &= -\alpha z - 4x - 4y - x^2. 
\end{align}
\end{subequations}
for $\alpha = 1.4$

\hspace{1cm}

\textbf{Chen:}
\begin{subequations}
\begin{align}
        \dot{x} &= a(y-x), \\
        \dot{y} &= (c-a)x - xz + cy,\\
        \dot{z} &= xy - bz, 
\end{align}
\end{subequations}
for $a = 40$, $b = 3$ and $c = 28$

\section{Supplementary figures}

This section presents additional figures supporting the main results.

\begin{figure}
    \includegraphics{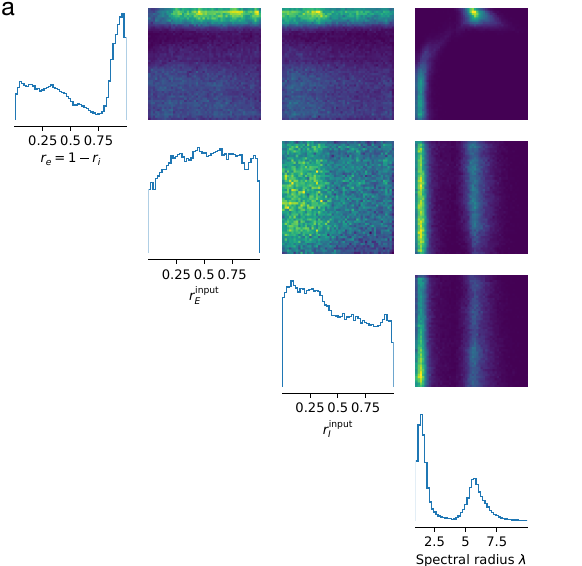}
    \caption{\textbf{Estimating optimal module-specific input connections in ESN that predicts four chaotic attractors.} Pairwise plot of marginal posterior distributions for four model parameters  $r_E$,$\lambda$,$r_i^{\textbf{in}}$,$r_e^{\textbf{in}}$) approximated with simulation-based inference. The plot is based on $10^5$ samples from the estimated posterior.}
    \label{fig:supp:sbi}
\end{figure}

\bibliographystyle{unsrt}
\bibliography{bibliography}